\documentclass{article}

\usepackage{arxiv}
\usepackage[utf8]{inputenc} % allow utf-8 input
\usepackage[T1]{fontenc}    % use 8-bit T1 fonts
\usepackage{hyperref}       % hyperlinks
\usepackage{url}            % simple URL typesetting
\usepackage{booktabs}       % professional-quality tables
\usepackage{amsfonts}       % blackboard math symbols
\usepackage{nicefrac}       % compact symbols for 1/2, etc.
\usepackage{microtype}      % microtypography
\usepackage{booktabs} % For formal tables
\usepackage{bbold}
\usepackage{amsmath}
\usepackage{pdfpages}
\usepackage{subfigure}
\usepackage{natbib}

\usepackage[ruled]{algorithm2e} % For algorithms

\SetAlFnt{\small}
\SetAlCapFnt{\small}
\SetAlCapNameFnt{\small}
\SetAlCapHSkip{0pt}
\IncMargin{-\parindent}
\newcommand{\todo}[1]{\textcolor{red}{#1}}

\title{Web Table Extraction, Retrieval and Augmentation: A Survey}

\author{
  Shuo Zhang\\
  University of Stavanger \\
  \texttt{shuo.zhang@uis.no} \\
   \And
 Krisztian Balog \\
  University of Stavanger\\
  \texttt{krisztian.balog@uis.no} \\
}

\begin{document}
\maketitle

\begin{abstract}
Tables are a powerful and popular tool for organizing and manipulating data.  A vast number of tables can be found on the Web, which represent a valuable knowledge resource.  The objective of this survey is to synthesize and present two decades of research on web tables.  In particular, we organize existing literature into six main categories of information access tasks: table extraction, table interpretation, table search, question answering, knowledge base augmentation, and table augmentation.  For each of these tasks, we identify and describe seminal approaches, present relevant resources, and point out interdependencies among the different tasks.
\end{abstract}

% keywords can be removed
\keywords{Table extraction \and table search \and table retrieval \and table mining \and table augmentation \and table interpretation}

\section{Introduction}

%\todo{Discuss ~\citep{Adelfio:2013:SET} somewhere!}

% Why we like tables
Tables are a practical and useful tool in many application scenarios. Tables can be effectively utilized for collecting and organizing information from multiple sources.  With the help of additional operations, such as sorting, filtering, and joins, this information can be turned into knowledge and, ultimately, can be used to support decision-making.  
Thanks to their convenience and utility, a large number of tables are being produced and are made available on the Web.  These tables represent a valuable resource and have been a focus of research for over two decades now.  In this survey paper, we provide a systematic overview of this body of research.

% How web tables differ from traditional tables and the different types of tables
Tables on the Web, referred to as \emph{web tables} further on in this paper, differ from traditional tables (that is, tables in relational databases and tables created in spreadsheet programs) in a number of ways.  First, web tables are embedded in webpages.  There is a lot of contextual information, such as the embedding page's title and link structure, the surrounding text, etc. that can be utilized. Second, web tables are rather heterogeneous regarding their quality, organization, and content.  For example, tables on the Web are often used for layout and navigation purposes.  
Among the different table types, \emph{relational tables} (also referred to as \emph{genuine tables}) are of special interest. These describe a set of entities (such as people, organizations, locations, etc.) along with their attributes~\citep{Wang:2002:DTH, cafarella:2008:URW, Crestan:2011:WTC,DEberiusBHTAL:2015:BDW,Zhang:2018:SES}.  Relational tables are considered to be of high quality, because of the relational knowledge contained in them.  However, unlike from tables in relational databases, these relationships are not made explicit in web tables; uncovering them is one of the main research challenges.

%similar to the relations, which are the tables stored in databases. In a database, a collection of relational data items is organized as a set of formally described tables. The relations are formally and consistently stored and keeping relationships with data stored in the same collection. Research on the relations includes relation search, relation ranking, column join, etc. However, tables on the Web are more arbitrary. To work with tables on the Web, it becomes more challenging. 

\begin{figure}[t]
   \centering
   \includegraphics[width=0.6\textwidth]{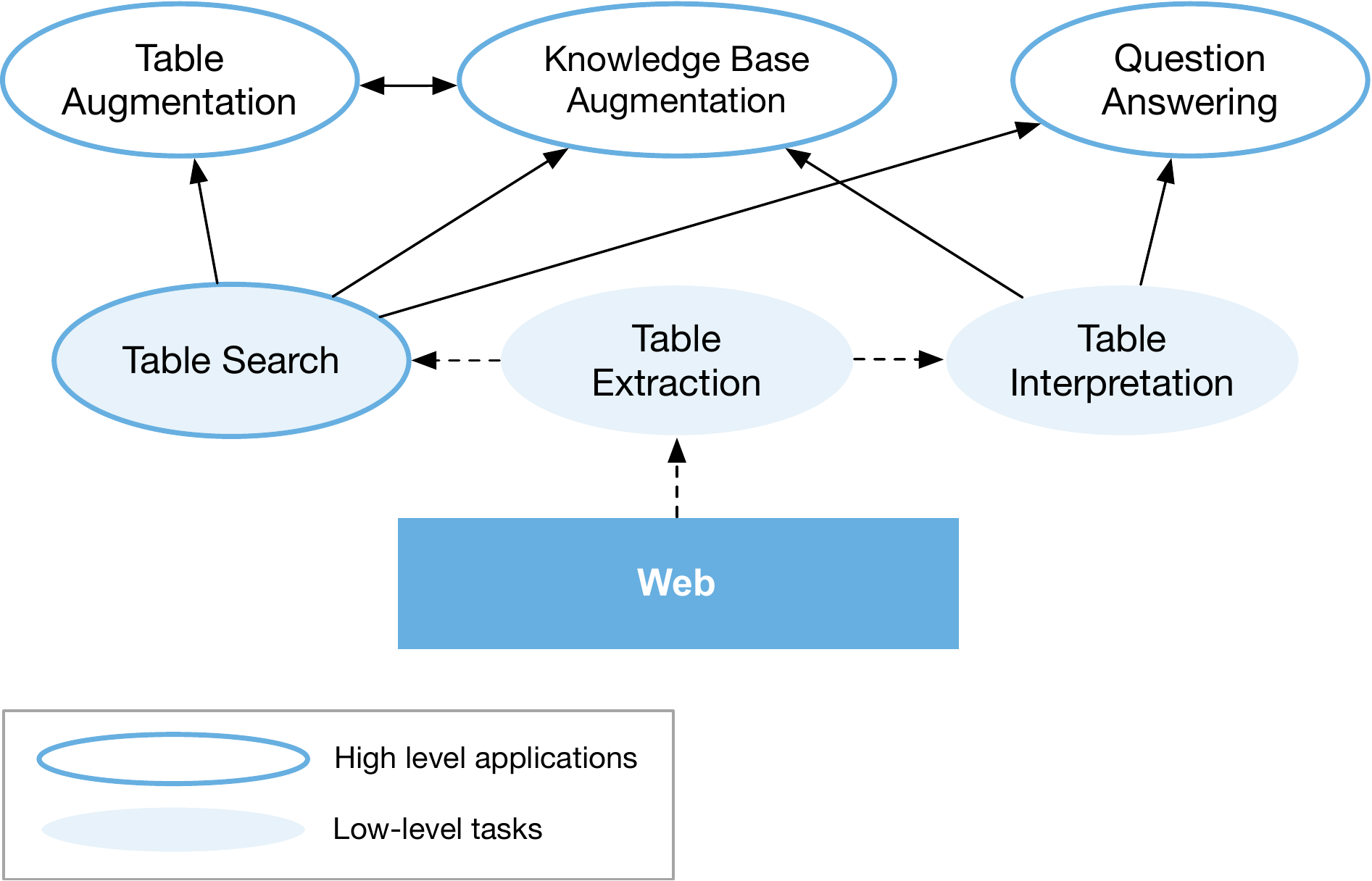} 
   \caption{Table-related information access tasks and their relationships.}
\label{fig:tasks}
\end{figure}

% Tasks
We organize relevant literature based on the task that is being addressed into six main categories. These are: table extraction (Sect.~\ref{sec:te}), table interpretation (Sect.~\ref{sec:interpret}), table search (Sect.~\ref{sec:tablesearch}), question answering on tables (Sect.~\ref{sec:qat}), knowledge base augmentation (Sect.~\ref{sec:kba}) and table augmentation (Sect.~\ref{sec:augment}).  
The relationship between the different tasks is shown in Fig.~\ref{fig:tasks}.

\begin{table}[t]
  \centering
%  \small
  \caption{Overview of table-related information access tasks.}
%  \vspace*{-0.5\baselineskip}
  \begin{tabular}{p{2.7cm}p{2cm}p{2.5cm}p{7cm}}
    \toprule
    \textbf{Task} & \textbf{Input} & \textbf{Output} & \textbf{Key references} \\
    \midrule
     Table extraction & Webpages & Tabular data 
     & \citet{Lehmberg:2016:LPC, Chen:2013:AWS, cafarella:2008:URW, Balakrishnan:2015:AWT, Cafarella:2009:DIR, DEberiusBHTAL:2015:BDW, Bhagavatula:2015:TEL} \\
	\hline
     Table interpretation & Table(s) & Structured data % (e.g., RDF triples) 
     & \citet{Wang:2002:MLB,Wang:2002:DTH,Lehmberg:2016:LPC,Chen:2013:AWS,cafarella:2008:URW, Crestan:2011:WTC, Lautert:2013:WTT, NishidaSHM:2017:USS, Venetis:2011:RST, Mulwad:2010:ULD, Fan:2014:AHM, Bhagavatula:2015:TEL, Wu:2016:ELW, Efthymiou:2017:MWT, Zhang:2013:MEA,HassanzadehWRS:2015:ULC, Mulwad:2013:SMP, Yoones:2014:KBA,Ibrahim:2016:MSE,Limaye:2010:ASW,Munoz:2014:ULD,Ritze:2016:PPW,RitzeB:2017:MWT} \\
    \hline
     Table search       & Query     & Ranked list of tables
     & \citet{Cafarella:2009:DIR, Pimplikar:2012:ATQ, Cafarella:2008:WEP,Bhagavatula:2013:MEM,AhmadovTELW:2015:THI,Lehmberg:2015:MSJ,DasSarma:2012:FRT,Yakout:2012:IEA,Nguyen:2015:RSS,Zhang:2018:AHT,Limaye:2010:ASW,Nargesian:2018:TUS,Zhang:2019:RRT}\\
    \hline
     Question Answering       & Natural language query & Structured data
     & \citet{PasupatL:2015:CSP, Sun:2016:TCS, Berant:2013:SPF,Fader:2014:OQA, Arvind:2015:NPI, Sarawagi:2014:OQQ, Banerjee:2009:LRQ} \\
     \hline
     Knowledge base augmentation & Table(s) & Structured data 
     & \citet{Bhagavatula:2015:TEL,Ritze:2015:MHT, RitzeB:2017:MWT, Ritze:2016:PPW, Lehmberg:2016:LPC,Ibrahim:2016:MSE,Zhang:2013:MEA,Yoones:2014:KBA,Fan:2014:AHM,Dong:2014:KVW} \\
    \hline
     Table augmentation & Table & Table 
     & \citet{DasSarma:2012:FRT, Yakout:2012:IEA, Cafarella:2008:WEP,Lehmberg:2015:MSJ,Bhagavatula:2013:MEM, Zhang:2017:ESA, Bhagavatula:2015:TEL,Venetis:2011:RST,Cafarella:2009:DIR, Zhang:2013:ISM,AhmadovTELW:2015:THI,Wang:2015:CEU,Zhang:2019:ADC} \\
    \bottomrule
  \end{tabular}
  \label{tbl:tasks}
%  \vspace*{-0.5\baselineskip}
\end{table}

\begin{itemize}
	\item \emph{Table extraction} refers to the process of detecting tables in webpages, extracting them, and storing them in a consistent format, resulting in a table corpus.  
	\item \emph{Table interpretation} aims to uncover the semantics of the data contained in a table, with the aim of making tabular data intelligently processable by machines.  This entails classifying tables according to some taxonomy, identifying what table columns are about, recognizing and disambiguating entity mentions in table cells, and uncovering the relationships between table columns.
	\item \emph{Table search} (or \emph{table retrieval}) is the task of answering a search query with a ranked list of tables.  The search query may be a sequence of keywords or it may be a table itself.
	\item \emph{Question Answering} utilizes structured data in tables for answering natural language questions. 
	\item \emph{Knowledge base augmentation} leverages tabular data for exploring, constructing, and augmenting knowledge bases. 
	\item \emph{Table augmentation} is directed at expanding an existing table with additional data. Specific subtasks include populating a table with new rows or columns, or finding missing cell values.
%	\item \emph{Other tasks}. \sz{Instead of searching tables in a table corpus, and returning an existing table as answer to a keyword search query, tables may also be constructed ``on the fly'' and presented as search results.  This scenario is motivated by cases where the desired table is not available in the corpus.  Tables can then be automatically generated from knowledge bases~\citep{Yang:2014:FPK, Zhang:2018:OTG}.}
\end{itemize}
For each of these tasks, summarized in Table~\ref{tbl:tasks}, we identify seminal work, describe the key ideas behind the proposed approaches, discuss relevant resources, and point out interdependencies among the different tasks.

The remainder of this paper is organized as follows. Next, in Sect.~\ref{sec:overview}, we introduce the different table types and table corpora.  Sections~\ref{sec:te}--\ref{sec:augment} are dedicated to the six main table tasks we have identified above. Finally, we conclude with a discussion of past progress and future research directions in Sect.~\ref{sec:concl}.

\section{Table Types and Corpora}
\label{sec:overview}

In this section, we formally introduce tables (Sect.~\ref{sec:overview:anatomy}), present various types of tables (Sect.~\ref{sec:overview:types}), and provide an overview %of table-related tasks (Sect.~\ref{sec:overview:tasks}) and 
of publicly available datasets (Sect.~\ref{sec:overview:corpora}).

\subsection{The Anatomy of a Table}
\label{sec:overview:anatomy}

\begin{figure}[t]
   \centering
   \includegraphics[width=0.7\textwidth]{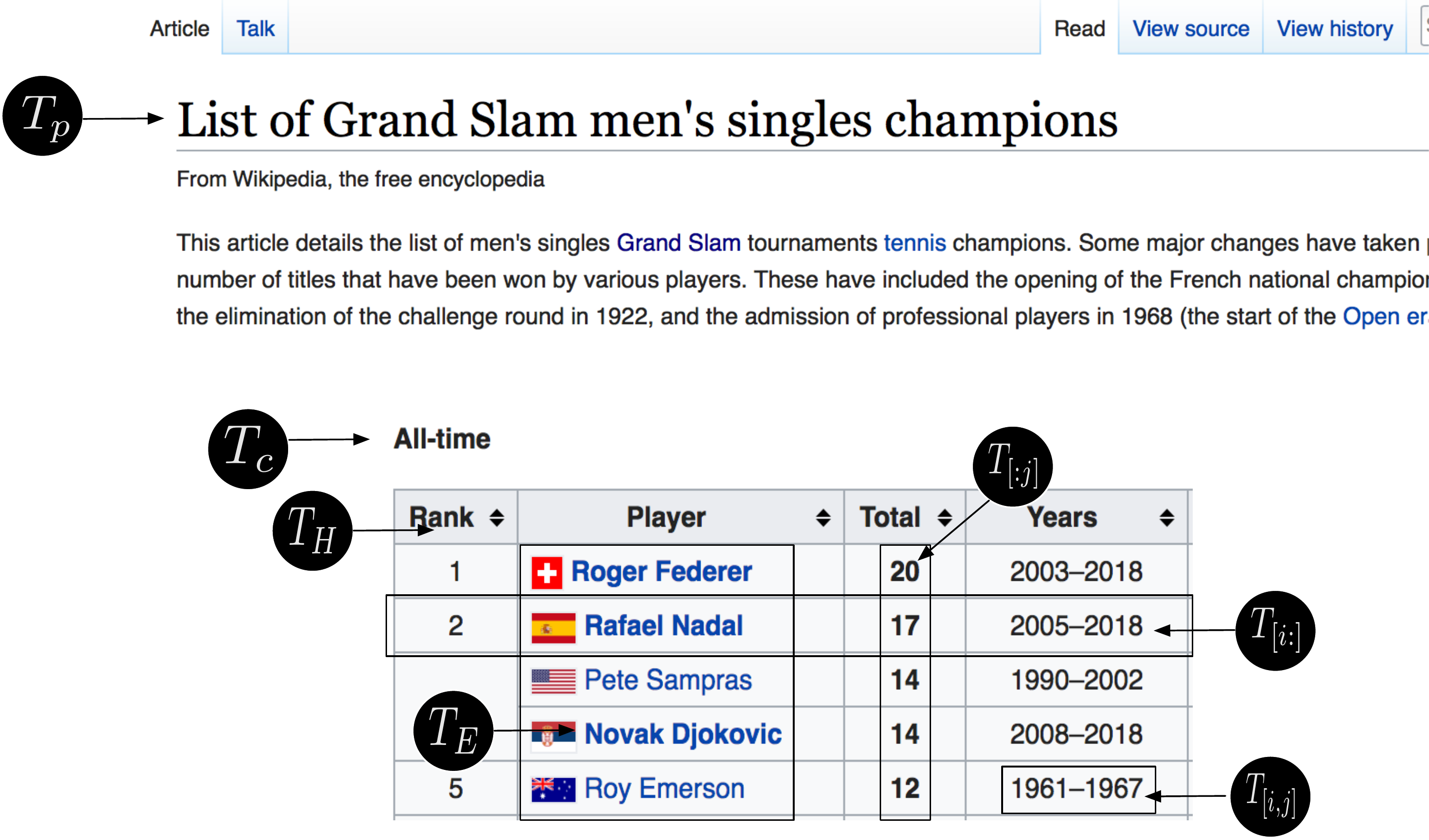} 
   \caption{Illustration of table elements in a web table: table page title ($T_p$), table caption ($T_c$), table headings ($T_H$), table cell ($T_{[i,j]}$), table row ($T_{[i,:]}$), table column ($T_{[:,j]}$), and table entities ($T_E$).}
\label{fig:table-element}
\end{figure}

A table $T$ is grid of cells arranged in rows and columns.  Tables are used as visual communication patterns, and as data arrangement and organization tools.
In this paper, our primary focus is on \emph{web tables}, that is, tables embedded in webpages.  Below, we define elements of a web table.  We refer to Fig.~\ref{fig:table-element} for an illustration. 

%``A table is an arrangement of data in rows and columns, or possibly in a more complex structure,'' as defined in Wikipedia.\footnote{\url{https://en.wikipedia.org/wiki/Table_(information)}}  % use @misc{}, url, title of article, date accessed

%\todo{@Shuo Add an example (relational) table here, showing the different table elements. It can be from Wikipedia. Table~\ref{tbl:notation} can then be removed.}

%

\begin{description}
	\item[Table page title] The table page title $T_p$ is the title of the webpage which embeds the table $T$.

	\item[Table caption] The caption of a table, $T_c$, is a short textual label summarizing what the table is about. 

	\item[Table headings] Table headings, $T_H$, is a list of labels defining what each table row/column is about.  Headings are typically in the first row/column in a table. 
	In case of relational tables (see below, in Sect.~\ref{sec:overview:types}), table headings are also referred to as \emph{table schema} or \emph{attribute names}.

	\item[Table cell] A table cell $T_{[i,j]}$ is specified with the row index $i$ and column index $j$. Table cells hold (possibly empty) values and are considered as atomic units in a table.
	
	\item[Table row] A table row $T_{[i,:]}$ is a list of table cells lying horizontally in line $i$ of a table.

	\item[Table column] A table row $T_{[:,j]}$ is a list of table cells lying vertically in column $j$ of a table.

	\item[Table entities] Tables often mention specific entities, such as persons, organizations, locations.  Table entities $T_E$ is a set consisting of all the entities that are mentioned in the table. 

\end{description}

%\begin{table}[t]
%  \centering
%  \caption{Notation used in this paper.}
%  \begin{tabular}{ll}
%    \toprule
%    \textbf{Symbol} & \textbf{Description} \\
%    \midrule
%    $T$ & Table \\ % & The entire table \\ %Data storage unit \\
%	$T_{[i,j]}$ & A specific table cell \\ % & A cell holding value \\
%    $T_{[i,:]}$ & A given table row \\ % & A horizontal collection of table cells \\
%    $T_{[:,j]}$ & A given table column \\ % & A vertical collection of table cells \\
%    
%%    $L$ & All column labels $L=(l_1, \dots, l_m)$ \\
%    $T_c$ & Table caption \\ %& Briefly summarizes the table's content \\
%    $T_p$ & The page title of web page embedding tables \\
%    
%    $T_E$ & The set of table entities \\
%    $T_H$ & The list of column headings (i.e., schema) \\ % labels & A list of column heading labels\\
%    \bottomrule
%  \end{tabular}
%  \label{tbl:notation}
%\end{table}

\subsection{Types of Tables}
\label{sec:overview:types}

A number of table classification schemes have been proposed in the literature. We start by reviewing those, then propose a normalized categorization based on the main aspects these share.

In early work, \citet{Wang:2002:DTH} make a distinction between genuine and non-genuine tables: 
\begin{itemize}
	\item \emph{Genuine tables} are leaf tables, i.e., do not contain other tables, lists, forms, images or other non-text formatting tags in a cell.  Furthermore, they contain multiple rows and columns.
	\item \emph{Non-genuines tables} refer to all those that are not leaf tables.
\end{itemize}

\noindent
\citet{cafarella:2008:URW} classify web tables into five main categories:  % parse 14.1B instances of HTML file with a table tag and 
\begin{itemize}
	\item \emph{Extremely small tables} are those having fewer than two rows or columns.
	\item \emph{HTML forms} are used for aligning form fields for user input.
	\item \emph{Calendars} are a specific table type, for rendering calendars.
	\item \emph{Non-relational tables} are characterized by low quality data, e.g., used only for layout purposes (many blank cells, simple lists, etc.).
	\item \emph{Relational tables} contain high-quality relational data.
\end{itemize}

\noindent
\citet{Crestan:2011:WTC} develop a fine-grained classification taxonomy, organized into a multi-layer hierarchy.
\begin{itemize}
	\item \emph{Relational knowledge tables} contain relational data.
	\begin{itemize}
		\item \emph{Listings} refer to tables consisting a series of entities with a single  attribute. In terms of layout direction, these are further classified as \emph{vertical listings} or \emph{horizontal listings}.
		\item \emph{Attribute/value tables} describe a certain entity along with its attributes.
		\item \emph{Matrix tables} have the same value type for each cell at the junction of a row and a column. \emph{Calendars}, for example, can be regarded as matrix tables. 
		\item \emph{Enumeration tables} list a series of objects that have the same ontological relation (e.g., hyponomys or siblings). 
		\item \emph{Form tables} are composed of input fields for the user to input or select values. 
	\end{itemize}
	\item \emph{Layout tables} do not contain any knowledge and are used merely for layout purposes.
	\begin{itemize}
		\item \emph{Navigational tables} are meant for navigating within or outside a website. 
		\item \emph{Formatting tables} are used for visually organizing content.
	\end{itemize}
\end{itemize}

\noindent
\citet{Lautert:2013:WTT} refine the classification scheme of \citet{Crestan:2011:WTC}.  %The first layer classification is almost the same as above.
\begin{itemize}
	\item \emph{Relational knowledge tables}
	\begin{itemize}
		\item \emph{Horizontal tables} place attribute names on top (column header). Each column corresponds to an attribute.
		\item \emph{Vertical tables} place attribute names on the left (row header). Each row represents an attribute.
		\item \emph{Matrix tables} are three dimensional data sets, where headers are both on the top and on the left.
	\end{itemize}
	\item \emph{Layout tables}, as before, are subdivided into \emph{navigational tables} and \emph{formatting tables}.
\end{itemize}

\noindent
Relational knowledge tables are further classified according to a secondary type taxonomy.
\begin{itemize}
	\item \emph{Concise tables} contain merged cells (i.e., cells with the same value conflated together) to avoid value repetition.
	\item \emph{Nested tables} contain a table in a cell.
	\item \emph{Multivalued tables} refer to tables containing multiple values in a single cell. If all values in one cell come from one domain, they are named as \emph{simple multivalued web tables}, if not, they are called \emph{composed multivalued value tables}.
	\item \emph{Splitted tables} present sequentially ordered repetitions in row/column headers (i.e., each label is repeated in every $x$ cell).
\end{itemize}

%Based on \cite{Lehmberg:2016:LPC, Crestan:2011:WTC}, \citet{NishidaSHM:2017:USS} classify genuine tables as relational tables, entity tables and matrix tables and other genuine tables as enumeration tables, calendar tables and form tables. Layout (non-genuine) tables have two categories: navigational tables and formatting tables.

\noindent
With a particular focus on web spreadsheets, \citet{Chen:2013:AWS} define the following type taxonomy:
\begin{itemize}
	\item \emph{Data frame spreadsheets} contain data frames, each consisting of two regions: data (numeric values) and headings (attribute names). These are further classified based on how they are arranged:
%have a two-part spreadsheet structured named data frame, which is composed of a block of numeric values and attribute on the top or left.
	\begin{itemize}
		\item \emph{Hierarchical left spreadsheets} place attributes on the left of the data region.
		\item \emph{Hierarchical top spreadsheets} put attributes on top of the data region.
	\end{itemize}
	\item \emph{Non-data frame (flat) spreadsheets} do not contain a data frame.
	\begin{itemize}
		\item \emph{Relation spreadsheets} can be converted into the relational model~\citep{Codd:1970:RMD}. %\todo{I suppose it refers to the relational model by Edgar F. Codd, 1969. But then add that paper as a reference.} % See: https://dl.acm.org/citation.cfm?doid=362384.362685
		\item \emph{Form spreadsheets} are designed for human-computer interaction.
		\item \emph{Diagram spreadsheets} are for visualization purposes.
		\item \emph{List spreadsheets} consist of non-numeric tuples.
		\item \emph{Other spreadsheets} include schedules, syllabi, scorecards, and other files without a clear purpose.
	\end{itemize}
\end{itemize}

\noindent
\citet{DEberiusBHTAL:2015:BDW} distinguish tables along two dimensions: content and layout.  
In terms of content, they adopt the classification scheme by \citet{Wang:2002:DTH}.  
Considering layout purposes, they sort tables according to their logical structure into the following categories:
\begin{itemize}
	\item \emph{Horizontal listings} align cells horizontally.
	\item \emph{Vertical listings} align cells vertically.
	\item \emph{Matrix tables} refer to numerical tables.
\end{itemize}

\noindent
\citet{Lehmberg:2016:LPC} distinguish between three main types of tables:
\begin{itemize}
	\item \emph{Relational tables} contain a set of entities, which could exist in rows (\emph{horizontal}) or columns (\emph{vertical}); the remainder of the cells contain their descriptive attributes.
	\item \emph{Entity tables} describe a certain entity.
	\item \emph{Matrix tables} refer to tables with numerical values only.
\end{itemize}

\noindent
The above categorization systems are quite diverse, which is not surprising considering that each was designed with a different use-case in mind.  
Nevertheless, we can observe two main dimensions along which tables are distinguished: content and layout. 
We propose a normalized classification scheme, which is presented to Table~\ref{tbl:types}.  In the remainder of this paper, we shall follow this classification when referring to a certain type of table. Among all table types, relational tables have received the bulk of attention in the literature.  Accordingly, we focus primarily on relational tables and the tasks based on them in this survey.

% NOTE: This is repetitive, doesn't help with synthesizing. 
%
%\sz{
%Binary classification (two categories) is to distinguish if the web tables have useful knowledge and the useful tables are then extracted from sources data, e.g., the scheme in~\cite{Wang:2002:DTH}.
%The first layer of most of the multi-layer classifications serves for the same purpose, e.g., the taxonomies in~\cite{Crestan:2011:WTC, Lautert:2013:WTT}
% for identifying useful web tables and the schema in~\citep{Chen:2013:AWS} for locating the tabular data area in spreadsheets. 
%The second layer of~\cite{Crestan:2011:WTC} has 5 categories for relational tables and 2 categories for layout tables, and that of~\cite{Lautert:2013:WTT} has 3 more categories for relational tables. The second layer of~\cite{Chen:2013:AWS} has 7 categories for utilizing spreadsheet data.
%Single-layer but multi-category taxonomy is designed either for recovering tabular knowledge, e.g. 5 categories in~\cite{cafarella:2008:URW} or constructing a classified web table corpus~\cite{DEberiusBHTAL:2015:BDW} by 3 types~\cite{Lehmberg:2016:LPC}. }

\begin{table}[t]
  \centering
  \caption{Classification of table types in this paper.  Our primary focus is on relational tables.}
  \begin{tabular}{lll}
    \toprule
    \textbf{Dimension} & \textbf{Type} & \textbf{Description} \\
    \midrule
    Content & Relational & Describes a set of entities with their attributes \\
            & Entity     & Describes a specific entity \\
            & Matrix     & A three dimensional data set, with row and column headers \\
            & Other      & Special-purpose tables, including lists, calendars, forms, etc. \\
    \midrule
    Layout  & Navigational & Tables for navigational purposes \\
            & Formatting & Tables for visual organization of elements \\
    \bottomrule
  \end{tabular}
  \label{tbl:types}
\end{table}

%
\if 0  % integrated into intro
%
\subsection{Overview of Tasks}
\label{sec:overview:tasks}

Below, we provide a brief description of various table-related information access tasks that we will be focusing on in this survey.

\begin{itemize}
	\item \emph{Table extraction} refers to the process of extracting tabular data from webpages and storing them consistent format, resulting in a table corpus.  
	\item \emph{Table interpretation} aims to uncover the semantics of the data contained in a table, thereby making it ``machine understandable.'' 
	\item \emph{Table search} (or \emph{table retrieval}) is the task of answering a search query with a ranked list of tables.  
	\item \emph{Table mining} focuses on discovering patterns from tables, which can subsequently be used for supporting various tasks, such as question answering, knowledge exploration, or knowledge base population. 
	\item \emph{Table augmentation} is directed at expanding an existing table with additional data. For instance, populating rows or columns, or finding missing cell values.
\end{itemize}
%
\fi
%

\subsection{Table Corpora}
\label{sec:overview:corpora}

A number of table corpora have been developed in prior work, which are summarized in Table~\ref{tbl:corpus}.  

\begin{table}[t]
  \centering
%  \small
  \caption{Overview of table corpora.}
  \begin{tabular}{p{4.2cm}p{2.5cm}p{1.5cm}p{4cm}}
    \toprule
     \textbf{Table corpora} & \textbf{Type} & \textbf{\#tables} & \textbf{Source} \\
    \midrule
    WDC 2012 Web Table Corpus & Web tables & 147M & Web crawl (Common Crawl) \\ 
    WDC 2015 Web Table Corpus & Web tables & 233M & Web crawl (Common Crawl) \\ 
    Dresden Web Tables Corpus & Web tables & 174M & Web crawl (Common Crawl) \\
    WebTables & Web tables & 154M & Web crawl (proprietary) \\
    WikiTables & Wikipedia tables & 1.6M & Wikipedia \\ 
    TableArXiv & Scientific tables & 0.34M & arxiv.org \\
    \bottomrule
  \end{tabular}
  \label{tbl:corpus}
\end{table}

\subsubsection{WDC Web Table Corpus}
\label{sec:overview:wdc}

There are two versions of WDC Web Table Corpus,\footnote{\url{http://webdatacommons.org/framework/}} which were released in 2012 and 2015, respectively.  The 2012 version contains 147 million web tables, which were extracted from the 2012 Common Crawl corpus (consisting of 3.5 billion HTML pages).  Tables in this corpus are roughly classified as relational or non-relational in terms of layout.  Statistically, 3.3 billion HTML pages were parsed and 11.2 billion tables were identified; tables that are not innermost (that is, contain other tables in their cells) were discarded.  1.3\% of the remaining tables (originating from 101 million different webpages) were labeled as relational tables.  Tables in this corpus are not classified further and neither is table context data provided.

The WDC 2015 Web Table Corpus, constructed by \citet{Lehmberg:2016:LPC}, contains 10.24 billion \emph{genuine} tables. The extraction process consists of two steps: table detection and table classification. The percentages of \emph{relational}, \emph{entity}, and \emph{matrix}  tables are 0.9\%, 1.4\%, and 0.03\%, respectively. The remaining 97.75\% accounts for \emph{layout} tables. 
%\sz{The WDC 2015 Web Table Corpus contains 233 million filtered tables.}
When storing a table, its orientation is also detected, indicating how the attributes are placed. In horizontal tables, the attributes are placed in columns, while in vertical tables they represent rows.
There are 90.26 million relational tables in total.
%A complete distribution of tables per top-level domain is provided. 62.25 million tables are extracted from the ``com'' domain, which is on top of the distribution. In the distribution of domains in Crawl, ``cappex.com'' is on top. 
Among those, 84.78 million are horizontal and 5.48 million are vertical. The average number of columns and rows in horizontal tables are 5.2 and 14.45. In vertical tables, these numbers are 8.44 and 3.66, respectively. \citet{Lehmberg:2016:LPC} also extract the column headers and classify each table column as being numeric, string, data, link, boolean, or list. The percentages of the numeric and string columns are 51.4\% and 47.3\%, respectively. Besides, the text surrounding the table (before and after) is also provided.
%5.47 million headers are identical from the total 462 million headers. Among these, empty string is the most popular used header. Based on the data type, \citet{Lehmberg:2016:LPC} 
%\sz{There are 139 million entity tables, 3 million matrix tables, and the rest are layout tables.  Additionally, numerous metadata fields are extracted.}
%
Furthermore, \citet{Lehmberg:2016:LPC} provide the English-language Relational Subset, comprising of relational tables that are classified as being in English, using a naive Bayesian language detector. The language filter considers a table's page title, table header, as well as the text surrounding the table to classify it as English or non-English. %The top-level domain and domains in Crawl distributions are similar to relational tables. 
The average number of columns and rows in this subset are 5.22 and 16.06 for horizontal tables, and 8.47 and 4.47 for vertical tables. The percentages of numeric and string columns are 51.8\% and 46.9\%.

A total of 139 million tables in the WDC 2015 Web Table Corpus are classified as entity tables. %62 million entity tables extracted from the ``com'' domain, which is also the most popular top-level domain. 
Out of these, 76.70 million are horizontal and 62.99 million are vertical tables. The average number of columns and rows are 2.40 and 9.08 for horizontal tables, and 7.53 and 2.06 for vertical tables. %Among over 100 million column headers, 17.81 million of them are identical. 
The column data types are quite different from that of relational tables. String columns are the most popular, amounting to 86.7\% of all columns, while numeric columns account for only 9.7\%.

The complete corpus as well as the different subcorpora are made publicly available.\footnote{\url{http://webdatacommons.org/webtables/\#results-2015}}

\subsubsection{Dresden Web Table Corpus}
\label{sec:overview:dresden}

\citet{DEberiusBHTAL:2015:BDW} also extracted tables from the Common Crawl web corpus.  The total number of tables is 174 million, which is reduced to 125 million after filtering with regards to content-based duplication.  The Dresden Web Table Corpus contains only the core table data, and not the entire HTML page. Even though the corpus is not available for download directly, the table extraction framework (extractor code and companion library for working with the data set) is made publicly available.\footnote{\url{https://wwwdb.inf.tu-dresden.de/misc/dwtc/}}

\subsubsection{WebTables}
\label{sec:overview:webtbl}

\citet{cafarella:2008:URW} extracted 154 million high-quality relational web tables from a (proprietary) general-purpose web crawl. Unfortunately, this corpus is not made public. However, frequency statistics of attributes, known as the ACSDb dataset (cf. Sec.~\ref{sec:augment:ce}), is available for download.\footnote{\url{https://web.eecs.umich.edu/~michjc/data/acsdb.html}}

\subsubsection{Wikipedia Tables}
\label{sec:overview:wikipedia}

\citet{Bhagavatula:2015:TEL} focused on Wikipedia and extracted 1.6 million high-quality relational tables.  Each table is stored as a JSON file, including table body, table caption, page title, column headers, and the number of row and columns. 
% ``data,'' ``title,'' ``caption,'' ``numCols,'' ``numericColumns'', ``pgTitle'', ``numDataRows'', etc. The ``data'' corresponds to the table body, where data is storied by table row. ``title'' is the column header, ``caption'' is table caption and ``pgTitle'' is the Wikipedia pages title. The rest are providing the statistics like number of numeric columns, number of columns, etc. 
The existing links in the tables are also extracted and stored in a separate file.  The corpus is available for download.\footnote{\url{http://websail-fe.cs.northwestern.edu/TabEL/}}

\subsubsection{Scientific Tables}

Scientific tables are a particular type of table, which contain valuable knowledge and are available in large quantities.
The TableArXiv corpus\footnote{\url{http://boston.lti.cs.cmu.edu/eager/table-arxiv/}} consists of 341,573 tables, extracted from physics e-prints on arxiv.org.  Along with the corpus, 105 information needs and corresponding relevance judgements are also provided for the task of scientific table search.

\section{Table Extraction}
\label{sec:te}

A vast number of tables can be found on the Web, produced for various purposes and storing an abundance of information.  These tables are available in heterogenous format, from HTML tables embedded in webpages to files created by spreadsheet programs (e.g., Microsoft Excel).  To conveniently utilize these resources, tabular data should be extracted, classified, and stored in a consistent format, resulting ultimately in a table corpus.  This process is referred to as \emph{table extraction}.
In this section, we present approaches for the table extraction task, organized around three main types of tables: web tables, Wikipedia tables, and spreadsheets.

\subsection{Web Table Extraction}
\label{sub:tecc}
Table extraction is concerned with the problem of identifying and classifying tables in webpages, which encompasses a range of more specific tasks, such as relational table classification, header detection, and table type classification.
These three tasks (relational table classification, header detection, and table type classification) are commonly approached as a supervised learning problem and employ similar features; these features are summarized in Tables~\ref{tbl:features1} and~\ref{tbl:features2}.
In Sects.~\ref{sub:tecc:rtc}--\ref{sub:tecc:ttc} we organize the literature according to the three tasks.
%\new{Therefore, we synthesize the literature based on these tasks: relational table classification in Sect.~\ref{sub:tecc:rtc}, header detection in Sect.~\ref{sub:tecc:hd}, and table type classification in Sect.~\ref{sub:tecc:ttc}.}

\subsubsection{Relational table classification}
\label{sub:tecc:rtc}
The identification of tables on the Web is usually straightforward based on HTML markup.  Tables, however, are also used extensively for formatting and layout purposes.  Therefore, web table extraction involves a data cleaning subtask, i.e., identifying and filtering out ``bad'' tables (where ``bad'' usually denotes non-relational tables).
\emph{Relational table classification} (also known as identifying high-quality or genuine tables) refers to the task of predicting whether a web table contains relational data.

One of the pioneering works utilizing tables on the Web is the WebTables project~\citep{Cafarella:2008:WEP,cafarella:2008:URW}.  \citet{cafarella:2008:URW} regard relational tables as high-quality tables, and filter those by training a rule-based classifier.  
The classifier uses table characteristics, like table size and table tags, as features.  The model is trained on a set of manually annotated tables (as being relational or non-relational) by two human judges. 
%Table ~\ref{tbl:features1} \sz{and Table ~\ref{tbl:features2}} list a set of selected features used for this \emph{relational table classification} task.  
As a result, they construct a high-quality table corpus, consisting of 154 million tables, filtered from 14.1 billion HTML tables (cf. Sect.~\ref{sec:overview:webtbl}).
\citet{Balakrishnan:2015:AWT} follow a similar approach for relational table classification, but use a richer set of features, which include both syntactic and semantic information. 
Syntactic features are related to the structure of the table, as in~\citep{cafarella:2008:URW} (e.g., number of rows and columns).
Semantic features are obtained by (1) determining whether the table falls into a boilerplate section of the containing page, (2) detecting subject columns (using a binary SVM classifier trained based on one thousand manually labeled tables), (3) identifying column types (which will be detailed later, in Sect.~\ref{sec:interpret:utc:cti}), (4) and detecting binary relationships between columns (by analyzing how these relationships are expressed in the text surrounding the table). 
\citet{Wang:2002:MLB} define a table as \emph{genuine}, if it is a leaf table where no subtable exists in any of the cells.  They employ machine learned classifiers (decision trees and support vector machines) to classify relational tables, using three main groups of features: layout features, content type features, and word group features.  The layout features and most of the content features are listed in Tables~\ref{tbl:features1} and~\ref{tbl:features2}.
As for word group features, \citet{Wang:2002:MLB} treat each table as a document and compute word frequency statistics.
%\sz{ further consider the word group features by .  are the statistics of the infrequent words in each table, and the word set are created in~\cite{Wang:2002:MLB}.}
%, features like are employed for training a tree classifier. In the second method,  They identify layout types for the analysis of a table's logic structure. 
In follow-up work, the authors also experiment with other machine learning methods (Naive Bayes and weighted kNN), using the same set of features~\citep{Wang:2002:DTH}.
Building on \citep{Wang:2002:MLB}, \citet{DEberiusBHTAL:2015:BDW} carry out relational table classification as well as classification according to layout type (vertical listings, horizontal listing, and matrix tables).  Their first method performs classification along both dimensions simultaneously, using a single layer.  Their second approach separates the two tasks into two layers, where the first layer executes table detection and, subsequently, the second layer determines the layout type.  Various machine learning methods are employed, including decision trees, Random Forests, and SVMs, using a combination of global and local features; a selection of features are listed in Table~\ref{tbl:features2}. 
As a result, \citet{DEberiusBHTAL:2015:BDW} classify millions of tables and generate the Dresden Web Table Corpus (DWTC, cf. Sect.~\ref{sec:overview:dresden}). 

To obtain metadata for relational tables, \citet{DEberiusBHTAL:2015:BDW} consider whether tables have a header row or not.  They find that 71\% of the tables in the corpus have a relational header.  For the remaining 29\%, they attempt to generate synthetic labels by comparing the column content to similar columns that have proper labels.
\citet{Cafarella:2009:DIR} design a system called OCTOPUS, which combines search, extraction, data cleaning, and integration.  %In OCTOPUS, tables are not limited to relations, e.g., HTML lists are also regarded as reliable. %, thus a larger corpus is used.
Further challenges related applying WebTables in practice, including table identification and table semantics recovery, are detailed in \citep{Balakrishnan:2015:AWT}.  The resulting system, Google Fusion Tables, is made publicly available.\footnote{\url{https://research.google.com/tables}}

\subsubsection{Header detection}
\label{sub:tecc:hd}
To extract data in a structured format, the semantics of tables need to be uncovered to some extent.  One question of particular importance is whether the table contains a header row or column.  This is known as the task of \emph{header detection}.  Headers may be seen as a particular kind of table metadata.  Header detection is commonly addressed along with the other two tasks and uses similar features (cf. Tables~\ref{tbl:features1} and~\ref{tbl:features2}).

\subsubsection{Table type classification} 
\label{sub:tecc:ttc}
Another type of metadata that can help to uncover table semantics is table type. \emph{Table type classification} is the task of classifying tables according to a pre-defined type taxonomy (cf. Sect.~\ref{sec:overview:types} for the discussion of various classification schemes). 
Additional metadata extracted for tables includes the embedding page's title, the table's caption, and the text surrounding the table.

%\todo{This is the part where we should introduce the other tasks: \emph{header detection} (which is one particular instance of table metadata extraction) and \emph{relational table classification}. Then, we say that similar features are used for all three (which are summarized in Table 4). Then, we should organize the discussion of works by task and use the task names consistently. The proposed order is i) relational table classification (also known as identifying high-quality or genuine tables), ii) header detection (one type of metadata), iii) table type classification (another type of metadata); it also generalizes i).}

\begin{table*}[t]
\centering
\caption{Selected features for relational table classification (RTC), header detection (HD), and table type classification (TTC) (Part 1/2). }
%\vspace*{-0.75\baselineskip}
\begin{tabular}{p{2.4cm}p{5.3cm}p{1.5cm}p{4.5cm}}
	\toprule
	\textbf{Features} & \textbf{Explanation} & \textbf{Task} & \textbf{Source} \\
	\midrule
	\multicolumn{3}{l}{\textbf{Global layout features}} \\
	\midrule
	Max rows & Maximal number of cells per row & RTC, TTC & \cite{Crestan:2011:WTC,DEberiusBHTAL:2015:BDW}\\
	Max cols & Maximal number of cells per column & RTC, TTC & \cite{Crestan:2011:WTC,DEberiusBHTAL:2015:BDW}\\
	Max cell length & Maximal number of characters per cell & RTC, TTC & \cite{Crestan:2011:WTC,DEberiusBHTAL:2015:BDW} \\
	\#rows & Number of rows in the table & RTC,  HD & \cite{cafarella:2008:URW}\\
	\#cols & Number of columns in the table & RTC,  HD & \cite{cafarella:2008:URW}\\
	\%rows & Percentage of rows that are mostly NULL & RTC & \cite{cafarella:2008:URW} \\
	\#cols non-string & Number of columns with non-string data & RTC & \cite{cafarella:2008:URW}\\
	$\mu$ & Average length of cell strings & RTC & \cite{cafarella:2008:URW}\\
	$\delta$ & Standard deviation of cell string length & RTC & \cite{cafarella:2008:URW}\\
	$\frac{\mu}{\delta}$ & Cell string length & RTC & \cite{cafarella:2008:URW}\\
	\%length one & Percentage of columns with $|len(row_1) - \mu|>2\delta$ & HD & \cite{cafarella:2008:URW}\\
	\%length two & Percentage of columns with $\delta \le |len(row_1) - \mu|\le 2\delta$ & HD & \cite{cafarella:2008:URW}\\
	\%length three & Percentage of columns with $|len(row_1)- \mu|<\delta$ & HD & \cite{cafarella:2008:URW}\\
	Avg rows & Average number of cells across rows & RTC, TTC & \cite{DEberiusBHTAL:2015:BDW,Wang:2002:MLB}\\
	Avg cols & Average number of cells across columns & RTC, TTC & \cite{DEberiusBHTAL:2015:BDW,Wang:2002:MLB} \\
	Avg cell length & Average number of characters per cell & RTC, TTC & \cite{Crestan:2011:WTC,DEberiusBHTAL:2015:BDW,Wang:2002:MLB}\\
	
		\bottomrule
\end{tabular}
\label{tbl:features1}
\end{table*}
	
\begin{table*}[t]
\centering
\caption{Selected features for relational table classification (RTC), header detection (HD), and table type classification (TTC) (Part 2/2). }
%\vspace*{-0.75\baselineskip}
\begin{tabular}{p{3cm}p{5.3cm}p{1.5cm}p{4.5cm}}
	\toprule
	\textbf{Features} & \textbf{Explanation} & \textbf{Task} & \textbf{Source} \\
	\midrule
	\multicolumn{3}{l}{\textbf{Layout features}} \\
	\midrule
	Std dev rows & Standard dev. of the number of cells per row & RTC & \cite{DEberiusBHTAL:2015:BDW,Wang:2002:MLB}\\
	Std dev cols & Standard dev. of the number of cells per column & RTC & \cite{DEberiusBHTAL:2015:BDW,Wang:2002:MLB}\\	
	Std dev cell length & Standard dev. of the number of characters per cell & RTC & \cite{Crestan:2011:WTC,DEberiusBHTAL:2015:BDW,Wang:2002:MLB}\\
	
	Local length avg & Average size of cells in segment & RTC & \cite{Crestan:2011:WTC,DEberiusBHTAL:2015:BDW}\\
	Local length variance & Variance of size of cells in segment & RTC & \cite{Crestan:2011:WTC,DEberiusBHTAL:2015:BDW}\\	
	
	\midrule
	\multicolumn{3}{l}{\textbf{Content features}} \\
	\midrule
	\%body non-string & Percentage of non-string data in table body & HD & \cite{cafarella:2008:URW}\\
	\%header non-string & Percentage of non-string data in the first row & HD & \cite{cafarella:2008:URW}\\
	\%header punctuation & Percentage of columns with punctuation in the first row & HD & \cite{cafarella:2008:URW}\\	
	Local span ratio & Ratio of cells with a $\langle$span$\rangle$ tag & RTC, TTC & \cite{Crestan:2011:WTC,DEberiusBHTAL:2015:BDW} \\	
	Local ratio header & Cells containing a $\langle$th$\rangle$ tag & RTC, TTC & \cite{Crestan:2011:WTC,DEberiusBHTAL:2015:BDW} \\	
	Local ratio anchor & Cells containing an $\langle$a$\rangle$ tag & RTC, TTC  & \cite{Crestan:2011:WTC,DEberiusBHTAL:2015:BDW}\\	
	Local ratio input & Cells containing an $\langle$input$\rangle$ tag & RTC, TTC & \cite{Crestan:2011:WTC,DEberiusBHTAL:2015:BDW} \\
	Ratio img & Ratio of cells containing images & RTC, TTC & \cite{Crestan:2011:WTC, DEberiusBHTAL:2015:BDW,Wang:2002:MLB} \\
	Ratio form & Ratio of cells containing forms & RTC, TTC & \cite{DEberiusBHTAL:2015:BDW,Wang:2002:MLB} \\	
	Ratio hyperlink & Ratio of cells containing hyperlinks & RTC, TTC & \cite{DEberiusBHTAL:2015:BDW,Wang:2002:MLB}\\
	Ratio alphabetic & Ratio of cells containing alphabetic characters & RTC, TTC & \cite{DEberiusBHTAL:2015:BDW,Wang:2002:MLB}\\	
	Ratio digit & Ratio of cells containing numeric characters & RTC, TTC & \cite{DEberiusBHTAL:2015:BDW,Wang:2002:MLB}\\	
	Ratio empty & Ratio of empty cells & RTC, TTC & \cite{DEberiusBHTAL:2015:BDW,Wang:2002:MLB}\\		
	Ratio other & Ratio of other cells & RTC, TTC & \cite{DEberiusBHTAL:2015:BDW,Wang:2002:MLB}\\
%	\midrule

		\bottomrule
\end{tabular}
\label{tbl:features2}
\end{table*}

The same features that are intended for relational table classification and header detection can also be used for table type classification~\cite{Wang:2002:MLB,Wang:2002:DTH,Lehmberg:2016:LPC,Chen:2013:AWS,cafarella:2008:URW}.  For example, the features listed in Tables~\ref{tbl:features1} and~\ref{tbl:features2} are used in \citep{DEberiusBHTAL:2015:BDW} for both relational table classification and table type classification. 
%\todo{$\Leftarrow$ The terminology in this part is confusing. Use the task names we introduced earlier.} 
Instead of directly classifying tables as relational or not, this can also be done indirectly by saying that a table is relational if relational information can successfully be extracted from it~\citep{Chen:2013:AWS}.
Table extraction is also involved in a number of other studies, but these datasets are not publicly available.  For example, with the purpose of data integration, \citet{Wang:2012:UTW} use a rule-based filtering method to construct a corpus of 1.95 billion tables.  For a type-classification study, \citet{Crestan:2011:WTC} extract a corpus of 8.2 billion tables.
Using a more fine-grained type taxonomy (see Sect.~\ref{sec:overview:types}), table type classification is approached as a multi-class classification problem.
\citet{Crestan:2011:WTC} propose a rich set of features, including global layout features, layout features, and content features. Global layout features include the maximum number of rows, cols, and maximum cell length. Layout features include average length of cells, length variance, and the ratio of row/column span. Content features include HTML features (based on HTML tags) and lexical features (based on cell content). 
As a follow-up work, \citet{Lautert:2013:WTT} additionally consider the category obtained in~\citep{Crestan:2011:WTC} as one features to further classify tables into a multi-layer taxonomy. The first layer of classification is similar to the one in \cite{Crestan:2011:WTC}. A second layer of classification focuses on relational knowledge, by additionally dividing relational knowledge tables into concise, nested, multivalued (simple or composed), and split tables.
\citet{Lehmberg:2016:LPC} construct a web table corpus from Common Crawl (WDC Web Table Corpus, cf. \ref{sec:overview:wdc}).  %They use a two step pipeline, consisting of filtering and classification steps.  
First, they filter out non-genuine tables (referred to as not innermost tables, i.e., tables that contain other tables in their cells) and tables that contain less than 2 columns or 3 rows.
Then, using the table extraction framework of DWTC, the filtered tables are classified as either relational, entity matrix, or layout tables~\citep{DEberiusBHTAL:2015:BDW}.
Recently, deep learning methods have also been used for table type classification. For example, \citet{NishidaSHM:2017:USS} regard a table as a matrix of texts, which is similar to an image.  Utilizing the type taxonomy from~\citep{Crestan:2011:WTC}, they design a framework named TabNet, consisting of RNN Encoder, CNN Encoder, and Classifier. The RNN Encoder encodes the input table cells to create a 3D table volume, like image data, in the first step. The CNN encoders encode the 3D table volume to capture table semantics, which is used for table type classification by the Classifier. Even though TabNet is designed to capture table structure, it can be applied to any matrix for type classification.

\subsection{Wikipedia Table Extraction}

Wikipedia tables may be regarded as a special case of web tables.  They are much more homogeneous than regular web tables and are generally of high quality.  Therefore, no additional data cleaning is required. \citet{Bhagavatula:2015:TEL} construct a Wikipedia table corpus, consisting of 1.6 million tables, with the objective of extracting machine-understandable knowledge from tables.  For details, we refer to Sect.~\ref{sec:overview:wikipedia}.

\subsection{Spreadsheet Extraction}

The Web contains a great variety and number of Microsoft Excel \emph{spreadsheets}. Spreadsheets are often roughly relational. %They are supposed to be well arranged in order to take advantage of the high value data. 
\citet{Chen:2013:AWS} design an automatic system to extract relational data, in order to support data integration operations, such as joins. A \emph{data frame} is defined as a block of numerical data.  \citet{Chen:2013:AWS} extract 410,554 Microsoft Excel files from the ClueWeb09 Web crawl by targeting Excel-style file endings that contain a data frame.  %A spreadsheet is said to be \emph{hierarchical} when it has a data frame.
Within a data frame, the attributes might lie on the left or top. \citet{Chen:2013:AWS} find that 50.5\% of the spreadsheets contain a data frame and 32.5\% of them have hierarchical top or left attributes (the rest are called \emph{flat} spreadsheets). Among the 49.5\% non-data frame spreadsheets, 22\% are relational, 10.5\% are forms, 3.5\% are diagrams, 3\% are lists, and 10.5\% are other spreadsheets.  For each spreadsheet, the extraction system firstly finds the data frame, then extracts the attribute hierarchy (top or left), and finally builds relational tuples (see Sect.~\ref{sec:interpret:re} for more details).

\section{Table Interpretation}
\label{sec:interpret}

\if 0
* 4.1.1:
   * add core column detection
   * how to deal with numeric and date columns
%* Add an illustration for Sec 4
%* Drop subsubsections and move them one level up
%* Drop "uncovering table semantics" subsection heading (but use this phrase in the intro of the section)
%* Put data translations under other tasks
\fi

\begin{figure}[t]
   \centering
   \includegraphics[width=1\textwidth]{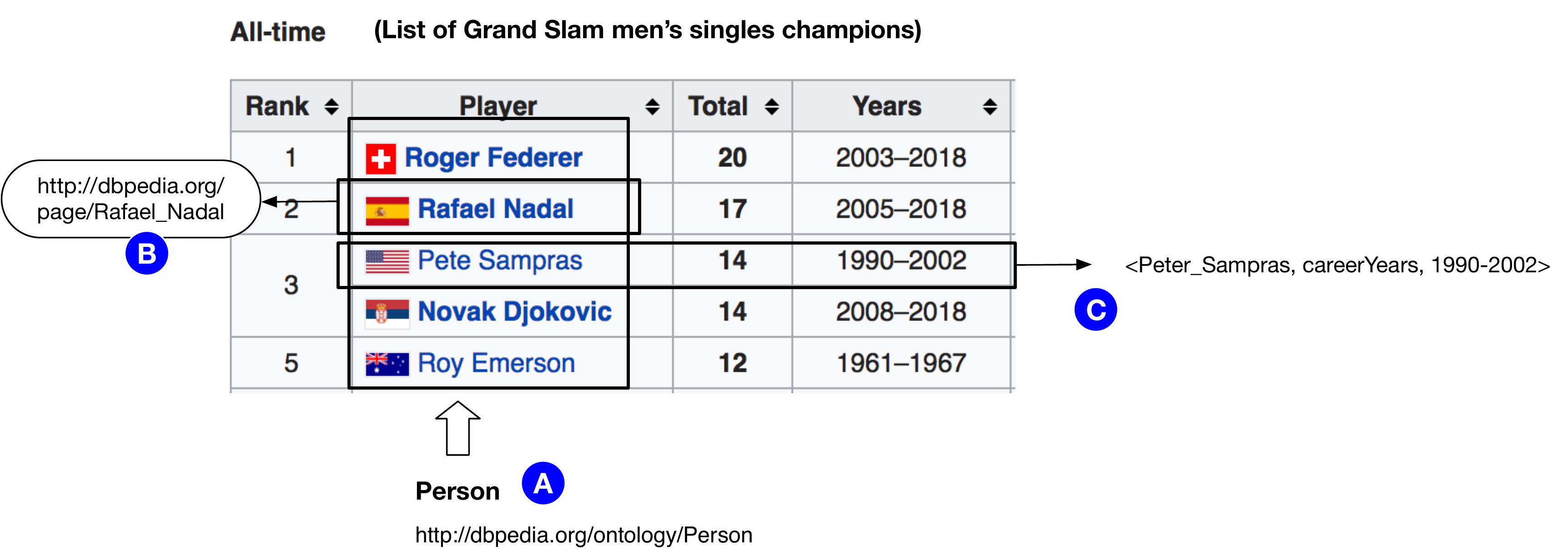} 
   \caption{Illustration of table interpretation: (A) Column Type Identification. (B) Entity Linking. (C) Relation extraction.}
\label{fig:tbl_int_illu}
\end{figure}

Table interpretation encompasses methods that aim to make tabular data processable by machines.  Specifically, it focuses on interpreting tables with the help of existing knowledge bases. 
\citet{Bhagavatula:2015:TEL} identify three main tasks aimed at uncovering table semantics: 
(1) \emph{column type identification}, that is, associating a table column with the type of entities or relations it contains,
(2) \emph{entity linking}, which is the task of identifying mentions of entities in cells and linking them to entries in a reference knowledge base, and
(3) \emph{relation extraction}, which is about associating a pair of columns in a table with the relation that holds between their contents.
See Figure~\ref{fig:tbl_int_illu} as the task illustration.
Table~\ref{tbl:interpret:tasks} provides an overview of studies addressing either or all of these tasks.
% These semantic annotations can then contribute to various table-related tasks, such as table search or combining tables via join or union~\citep{Venetis:2011:RST}.

%\subsection{Uncovering Table Semantics}
%\label{sec:interpret:utc}

%
%

%% OLD TABLE LAYOUT

%\begin{table}[t]
%  \centering
%  \small
%  \caption{Overview of table interpretation tasks.}
%  \begin{tabular}{p{3.7cm}p{7.7cm}}
%    \toprule
%    \textbf{Task} & \textbf{Key references} \\
%    \midrule
%     Column Type Identification & \citet{Venetis:2011:RST,Mulwad:2010:ULD,Fan:2014:AHM,Wang:2012:UTW,Lehmberg:2016:WTC,Zhang:2017:EAE,Zhang:2013:ISM,Limaye:2010:ASW} \\
%	\hline
%     Entity Linking & \citet{Limaye:2010:ASW,Bhagavatula:2015:TEL,Wu:2016:ELW,Efthymiou:2017:MWT,Zhang:2017:EAE,Mulwad:2010:ULD,Ibrahim:2016:MSE,Zhang:2013:MEA,HassanzadehWRS:2015:ULC,RitzeB:2017:MWT,Ritze:2016:PPW} \\
%     \hline
%     Relation Extraction & \citet{Venetis:2011:RST,Mulwad:2010:ULD,Mulwad:2013:SMP,Zhang:2017:EAE,Yoones:2014:KBA,Munoz:2014:ULD,Zwicklbauer:2013:TDW,Chen:2013:AWS}\\
%    \bottomrule
%  \end{tabular}
%  \label{tbl:interpret:tasks}
%\end{table}

\begin{table}[t]
  \centering
%  \small
  \caption{Overview of table interpretation tasks addressed in various studies.}
%{\color{blue}
  \begin{tabular}{lccc}
    \toprule
    \textbf{Reference} 
    	& \textbf{Column type}
    	& \textbf{Entity}
    	& \textbf{Relation} \\
    	& \textbf{Identification}
    	& \textbf{Linking}
    	& \textbf{Extraction} \\
    \midrule
    \citet{Bhagavatula:2015:TEL} & & \checkmark & \\
    \citet{Chen:2013:AWS} & & & \checkmark \\
    \citet{Efthymiou:2017:MWT} & & \checkmark & \\
    \citet{Fan:2014:AHM} & \checkmark & & \\
    \citet{HassanzadehWRS:2015:ULC} & & \checkmark & \\
    \citet{Ibrahim:2016:MSE} & & \checkmark & \\
    \citet{Lehmberg:2016:WTC} & \checkmark & & \\
    \citet{Limaye:2010:ASW} & \checkmark & \checkmark & \\
    \citet{Munoz:2014:ULD} & & & \checkmark \\
    \citet{Mulwad:2013:SMP} & & & \checkmark \\
    \citet{Mulwad:2010:ULD} & \checkmark & \checkmark & \checkmark \\
    \citet{RitzeB:2017:MWT} & & \checkmark & \\
    \citet{Ritze:2016:PPW} & & \checkmark & \\
    \citet{Yoones:2014:KBA} & & & \checkmark \\
    \citet{Venetis:2011:RST} & \checkmark &  & \checkmark \\
    \citet{Wang:2012:UTW} & \checkmark & & \\
    \citet{Wu:2016:ELW} & & \checkmark & \\
    \citet{Zhang:2013:ISM} & \checkmark & \checkmark & \\
    \citet{Zhang:2017:EAE} & \checkmark & \checkmark & \checkmark \\
%    \citet{Zwicklbauer:2013:TDW} & & & \checkmark \\
    \bottomrule
  \end{tabular}
%} % temp coloring 
  \label{tbl:interpret:tasks}
\end{table}

\subsection{Column Type Identification}
\label{sec:interpret:utc:cti}
\begin{table}[t]
  \centering
%  \small
  \caption{Comparison of column type identification tasks.}
  \small
%{\color{blue}
  \begin{tabular}{llll}
    \toprule
    \textbf{Reference} & \textbf{Knowledge base} & \textbf{Method} &  \\
    \midrule
     \citet{Fan:2014:AHM} & Freebase & Concept-based method + crowdsourcing \\
     \citet{Lehmberg:2016:WTC} & DBpedia & Feature-based classification \\
     \citet{Mulwad:2010:ULD} & Wikitology & Entity search \\
     \citet{Wang:2012:UTW} & Probase & Heading-based search \\
     \citet{Venetis:2011:RST} & Automatically built IS-A database & Majority vote \\ 
     \citet{Zhang:2013:ISM} & - & Semantic graph method\\
     \citet{Zhang:2017:EAE} & Freebase & Unsupervised featured-based method \\
    \bottomrule
  \end{tabular}
%} % temp coloring   
  \label{tbl:cti:tasks}
\end{table}
In relational tables, the \emph{core column} (also referred to as \emph{subject column}, \emph{name column}, or \emph{entity column}~\citep{Lehmberg:2016:WTC}) is a special column that contains entities. Commonly, this is the leftmost column in a table (and other table columns correspond to attributes or relationships of these entities). 
The identification of the core column is a central pre-processing step for entity linking, table augmentation, and relation extraction. 
Most of the existing work assumes the presence of a single core column.  Such tables are also known as single-concept relational tables.
However, in some cases, a relational table might have multiple core columns that may be located at any position in the table~\citep{Braunschweig:2015:FWT}, called a multi-concept relational table. \citet{Braunschweig:2015:FWT} extend a single-concept method, which utilizes table headings as well as intrinsic data correlations, with more features, like the correlation with the left neighbor, to determine all the core columns.
We focus on single-concept relational tables in the remainder of this section.

Generally, \emph{column type identification} is concerned with determining the types of columns, including locating the core column.  This knowledge can then be used to help interpret a table. Table~\ref{tbl:cti:tasks} displays a summary of the methods, which we shall discuss below.
\citet{Venetis:2011:RST} argue that the meaning of web tables is ``only described in the text surrounding them. Header rows exist in few cases, and even when they do, the attribute names are typically useless''~\citep{Venetis:2011:RST}.  
Therefore, they add annotations to tables to describe the sets of entities in the table (i.e., column type identification).
This is accomplished by leveraging an IS-A database of entity-class pairs. 
This IS-A database is created by aggregating all the entity-class $\langle e,C \rangle$ pairs that are mined from the Web (100 million English documents using 50 million anonymized queries) using the pattern ``$C$ [such as|including] $e$ [and|,|.].''
A class label is assigned to a column if a certain fraction of entities in that column is identified with that label in the IS-A database.  
\citet{Venetis:2011:RST} conclude that using a knowledge base (YAGO) results in higher precision, while annotating against the IS-A database has better coverage, i.e., higher recall.
\citet{Mulwad:2010:ULD} map each cell's value in a column to a ranked list of classes, and then selects a single class which best describes the whole column. To get the ranked list of classes, a complex query, based on cell values, is submitted to the Wikitology knowledge base~\citep{Syed:2010:WNH}. Possible class labels are obtained by utilizing the relevant entities in the knowledge base. Then, a PageRank-based method is used to compute a score for the entities' classes, from which the one with the highest score is regarded as the class label. 
Mapping each column to one of the four types (``Person'', ``Place'', ``Organization,'' and ``Other''), \citet{Mulwad:2010:ULD} achieve great success on ``Person'' and ``Places,'' and moderate success on ``Organization'' and ``Other'' types, due to their sparseness in the reference knowledge base.

Because of the inherent semantic heterogeneity in web tables, not all tables can be matched to a knowledge base using pure machine learning methods. \citet{Fan:2014:AHM} propose a ``two-pronged'' approach for matching web tables' columns to a knowledge base. First, a concept-based method is used to map each column to the best knowledge base concept.  Specifically, they employ Freebase as the concept catalog. Second, a hybrid human-machine framework discerns the concepts for some exceptional columns manually. The matches between table columns and their candidate concepts are represented as a bipartite graph, where relationships correspond to edges. \citet{Fan:2014:AHM} employ crowdsourcing for this task, and find that a higher payment leads to better accuracy.

%\citer{Wang:2012:UTW}, 
%a table corpus consisting of 1.95 billion raw HTML tables is constructed. Useful tables are firstly filtered based on a rule-based method and then 
A table corpus is constructed in~\citep{Wang:2012:UTW} and it is classified according to a probabilistic taxonomy called Probase, which is able to understand entities, attributes, and cells in tables. 
%With the extracted knowledge, \citet{Wang:2012:UTW} implement a semantic search engine for keyword table search. 
To get the table semantics, a top-$k$ candidates concepts are returned based on the table headings, which is similar to the idea in~\cite{Limaye:2010:ASW} (cf. Sect.~\ref{sec:interpret:el}). The candidate concepts assist to detect entities in a given column by computing the maximum number of common concepts. In turn, the entity column type is obtained based on the confidence of the concepts. 
\citet{Wang:2012:UTW} demonstrate that table headers can help to understand the columns as well as to identify the core column.

\citet{Lehmberg:2016:WTC} propose a categorization scheme for web table columns that distinguishes the different types of relations that appear in tables on the Web. First, a binary relation is a relation that holds between the core column and the values in another column, e.g., populations of cities. Second, an N-ary relation is a relation that holds between the core column and additional entities and values in other columns. Third, an independent column is one that has no direct relation with the core column. \citet{Lehmberg:2016:WTC} propose a feature-based classifier that distinguishes between these three types of relations for better table interpretation.

\citet{Zhang:2017:EAE} presents TableMiner+ for semantic table interpretation, where core column detection and type identification linking are executed at the same stage. \citet{Zhang:2017:EAE} first simply uses regular expressions and classifies cells as ``empty,'' ``entities,'' ``numbers,'' ``data,'' ``text,'' or ``other.'' Then, evidence is gathered from the Web for each column to predict the likelihood of it being the subject (core) column.  Specifically, a keyword query is composed from all text content in each row, and the subject entity in this row is detected by recognizing the top-ranked page. Finally, an unsupervised feature-based method is employed to find the core column and type by aggregating evidence across all rows. Features include the fraction of empty cells, the fraction of cells with unique content, context match score (heading frequency within surrounding text), and web search score. 
%These features are normalized and fed into relative scores by maximizing the same feature type. The column with the highest score is taken as the core column.
The main differences between TableMiner+ and other methods are twofold: (1) TableMiner+ uses context outside the tables while others not, and (2) it adopts an iterative process to optimize and enforce the interdependence between different annotation tasks (entity linking and relation extraction).

The above methods work well for string values and static attributes but perform poorly for numeric and time-varying attributes. \citet{Zhang:2013:ISM} build a semantic graph over web tables suited for numeric and time-varying attributes by annotating columns with semantic labels, like timestamp, and converting columns by comparing with columns from other tables. While this method is designed for entity augmentation, it can also be utilized for column type identification. 

\subsection{Entity Linking}
\label{sec:interpret:el}
\begin{table}[t]
  \centering
%  \small
  \caption{Comparison of entity linking tasks.}
  \footnotesize
%  {\color{blue}
  \begin{tabular}{p{3.5cm}p{5cm}p{6cm}}
    \toprule
    \textbf{Reference} & \textbf{Knowledge base} & \textbf{Method} \\
    \midrule
    \citet{Bhagavatula:2015:TEL} & YAGO & Graphical model  \\
    \citet{Efthymiou:2017:MWT} & DBpedia & Vectorial representation and ontology matching \\
     \citet{HassanzadehWRS:2015:ULC} & DBpedia, Schema.org, YAGO, Wikidata, and Freebase & Ontology overlap$^a$ \\
     \citet{Ibrahim:2016:MSE} & YAGO & Probabilistic graphical model \\
     \citet{Lehmberg:2017:SWT} & DBpedia  & Feature-based method \\      
     \citet{Lehmberg:2016:LPC} & Google Knowledge Graph & - \\
    \citet{Limaye:2010:ASW} & YAGO catalog, DBpedia, and Wikipedia tables & Inference of five types of features$^b$ \\
    \citet{Mulwad:2010:ULD} & Wikitology & SVM classifier \\
     \citet{RitzeB:2017:MWT} & DBpedia  & Feature-based method \\
     \citet{Ritze:2015:MHT,Ritze:2016:PPW} & DBpedia  & Feature-based method \\
    \citet{Wu:2016:ELW} & Chinese Wikipedia, Baidu Baike, and Hudong Baike & Probabilistic method$^c$ \\
     \citet{Zhang:2013:MEA} & DBpedia & Instance-based schema mapping \\ 
    \citet{Zhang:2017:EAE} & Freebase & Optimization  \\
    \bottomrule
    \multicolumn{3}{l}{\footnotesize{$^a$ KB comparison}} \\
    \multicolumn{3}{l}{\footnotesize{$^b$ Designed for table search}} \\
    \multicolumn{3}{l}{\footnotesize{$^c$ Multiple KBs}} \\
  \end{tabular}
%  }
  \label{tbl:el:tasks}
\end{table}
Recognizing and disambiguating specific entities (such as persons, organizations, locations, etc.), a task commonly referred to as \emph{entity linking}, is a key step to uncovering semantics~\citep{Bhagavatula:2015:TEL}. Since many web tables are relational, describing entities, entity linking is a key step to understanding what the table is about.
A number of table-related tasks, such as row population~\citep{Zhang:2017:ESA, Wang:2015:CEU}, column population~\citep{Zhang:2017:ESA}, and table search~\citep{Zhang:2018:AHT}, rely on entity linking in tables.
Table~\ref{tbl:el:tasks} compares the tasks we will discuss below. 

\citet{Limaye:2010:ASW} pioneered research on table entity linking. They introduce and combine five features, namely, the TF-IDF scores between cell text and entity label, the TF-IDF scores between the column header and the type label, the compatibility between column type and cell entity, compatibility between relation and pair of column types, and the compatibility between relation and entity pairs. %The compatibility is formalized by fractions, IDF, etc., between them.
%\sz{These features are fed into the collective objectives named ``inference''. I.e., values of each features are assigned to their corresponding variables, which are combined as the objective function.} 
%Additionally, their goal is to answer queries over web tables. 
Their idea of a factor graph based entity linking approach influenced later research. For example,
\citet{Bhagavatula:2015:TEL} design a system called TabEL for table entity linking. 
TabEL employs a graphical model that ``assigns higher likelihood to sets of entities that tend to co-occur in Wikipedia documents and tables''~\citep{Bhagavatula:2015:TEL}. Specifically, it uses a supervised learning approach and annotated mentions in tables for training. TabEL focuses on Wikipedia table and executes mention identification for each table cell, then obtains a set of candidate entities for disambiguation. The disambiguation technique is based on the assumption that entities in a given row and column tend to be related. They use a collective classification technique to optimize a global coherence score for a set of entities in a given table.
By comparing against traditional entity linking methods for unstructured text, \citet{Bhagavatula:2015:TEL} demonstrate the need for entity linking methods designed specifically for tables.

Unlike most methods, which consider a single knowledge base, 
\citet{Wu:2016:ELW} propose an entity linking method for web tables that considers multiple knowledge bases to ensure good coverage. From each knowledge base, entities whose names share at least one word with the content of a given table cell are taken as candidates.  Then, an entity disambiguation graph is constructed, consisting of mention nodes, entity nodes, mention-entity edges, and entity-entity edges.  The method utilizes entity linking ``impact factors,'' which are two probabilities, for ranking candidates and for disambiguating entities, based on mention nodes and edges. To incorporate multiple knowledge bases, ``same-As'' relations between entities from different knowledge bases are leveraged to reduce errors and to improve coverage. 
This system shares many similarities with TabEL. TabEL, however, does not consider synonyms and deals with a single KB.
\citet{Efthymiou:2017:MWT} propose three unsupervised annotation methods for matching web tables with entities. The first is a lookup-based method, which relies on the minimal entity context from the tables to discover correspondences to the knowledge base. A second method exploits a vectorial representation of the rich entity context in a knowledge base to identity the most relevant subset of entities in web tables. The third method is based on ontology matching, and exploits schematic and instance information of entities available both in a knowledge base and in a web table.
\citet{Efthymiou:2017:MWT} find that hybrid methods that combine the second and third methods (in any order) tend to perform best.
The column type identification component of TableMiner+~\citep{Zhang:2017:EAE} has already been discussed earlier, in Sect.~\ref{sec:interpret:utc:cti}.  
Building on this, TableMiner+~ uses the partial annotations from column type identification for all columns to guide entity linking in the rest of the table. 
It re-ranks table rows under the assumption that some cells are easy to disambiguate, i.e., they have more candidates or the text is less ambiguous (candidate sampling). In each iteration of this so-called \emph{learning} phase,
it searches new candidates and compares the feature representation of each candidate entity (web search results) against all the feature representations of that cell (using the same features as for column type identification). The associated concepts with the highest scoring entity are gathered as candidate concepts for the column. These are further compared against those from the previous iteration in the \emph{learning} phase (optimization). The process is repeated until convergence is reached.
%interacts with candidate sampling and becomes stable once convergence.}  

\citet{Mulwad:2010:ULD} exploit the predicted class labels for columns (see Sect.~\ref{sec:interpret:utc:cti}) as additional evidence, to link entities in table cells.  A knowledge base is queried to construct a feature vector, which comprises the entity's retrieval score, Wikipedia page length, PageRank, etc., which are used for computing the similarity score against the table cell's value.
The feature vectors are input to an SVMRank classifier, which outputs a ranked list of entities. The top-ranked entity is selected and is used to introduce two more features for a final classification (the SVM rank score for the top-ranked entity and the score difference between the top two entities). The final classification yields a binary outcome whether the entity should be linked or not.  Similar to the column type identification task, this method performs very well on the ``Person'' and ``Place'' entity types, achieves moderate accuracy on ``Organization,'' and low accuracy on ``Other'' (for the same reason of sparseness, as before).
A similar approach is taken by \citet{Lehmberg:2016:LPC}, but they perform entity linking in table cells first, using the Google Knowledge Graph, and then use this information for getting class labels for columns. 

Another study on knowledge base matching in ~\cite{Ibrahim:2016:MSE} aims to overcome the problem of table matching and aggregation by making sense of entities and quantities in web tables. \citet{Ibrahim:2016:MSE} map the table elements of table headers, entity tables cells, and numeric table cells to different knowledge bases. Specifically, (1) tables headers denote classes or concepts and are linked to a taxonomic catalog or to Wikipedia pages, (2) named entities are mapped to a knowledge base (YAGO), and (3) numeric cells, which denote quantities, are mapped to normalized representations. An interesting observation made about quantity linking is that many of the linking errors are (1) due to the absence of specific measures or units and (2) because of ambiguous headings, like ``Nat.''

As mentioned in Sect.~\ref{sub:tecc}, a \emph{relational table} refers to an entity-attribute table, where a set of entities and their attributes are listed. \citet{Zhang:2013:MEA} propose an instance-based schema mapping method to map entity-attribute tables to a knowledge base. In \cite{Zhang:2013:MEA}, an entity-attribute table is supposed to have a key column, which contains a set of entities. Each tuple is an entity with its attributes. Then, memory-based indexes are used to judge whether a tuple contains candidate entities, resulting in an evidence mapping vector.  This vector is then used for finding a table-to-KB schema mapping, which essentially serves as a bridge between web tables and knowledge bases.
 
The choice of the knowledge base for uncovering table semantics is important. \citet{HassanzadehWRS:2015:ULC} give a detailed study on the utility of different knowledge bases, including DBpedia, Schema.org, YAGO, Wikidata, and Freebase.  The method of concept linking in \cite{HassanzadehWRS:2015:ULC} is tagging columns with entity types (classes) in the knowledge base. Specifically, they firstly get the basic statistical distribution of tables sizes and values. Then, with the help of the selected knowledge base, the distribution of overlap scores in the ontology is obtained.  Finally, these scores can give an indication of how well the table's content is covered by the given knowledge base. 

\citet{RitzeB:2017:MWT} study the utility of different features for entity linking in tables.
These features are extracted from the table itself (such as entity label, table, URL, page title, and surrounding text) or from the knowledge base (such as instance label and classes). 
They introduce a specific similarity linker for each feature, resulting in similarity matrices, representing feature-specific results.  These matrix predictors can be used to decide which features to use for which web table.
\citet{Ritze:2016:PPW} implement the T2K Match framework~\citep{Ritze:2015:MHT} to map the WDC Web corpus to DBpedia, for knowledge base extension (entity linking happens the same time with, and rely on, schema matching and table type identification). Taking table content as evidence, the incomplete and unclear values of DBpedia can be filled and corrected.
They find that ``only 1.3\% of all tables that were extracted from the Web crawl contained relational data. Out of these relational tables, about 3\% could be linked to DBpedia''~\citep{Ritze:2016:PPW}.
%\citet{Ritze:2016:PPW} further introduce three fusion strategies for deciding which value to use as output.  Their best method achieves an F1-score of 0.7.
However, the above methods tend to perform better for large tables, i.e., tables with several rows. It is considered as one of the main limitations of linking tabular mentions to DBpedia. To overcome this, \citet{Lehmberg:2017:SWT} stitch tables, i.e., merge tables from the same website as a single large table, in order to improve entity linking performance.

\subsection{Relation Extraction}
\label{sec:interpret:re}
\begin{table}[t]
  \centering
  \footnotesize
  \caption{Comparison of relation extraction tasks.}
  \small
%   {\color{blue}
  \begin{tabular}{p{4cm}p{2cm}p{2.8cm}p{4.5cm}}
    \toprule
    \textbf{Reference} & \textbf{Knowledge base} & \textbf{Method} & \textbf{Source of extraction} \\
    \midrule
      \citet{Chen:2013:AWS} & - & Classification & Each value in the value region \\
      \citet{Munoz:2014:ULD} & DBpedia & Look-up based & Any pair of entities in the same row\\
     \citet{Mulwad:2013:SMP} & DBpedia & Semantic passing & Any pair of columns \\
     \citet{Mulwad:2010:ULD} & DBpedia & Utilizing CTI and EL & Any pair of columns  \\
      \citet{Yoones:2014:KBA} & YAGO, PATTY & Probabilistic & Any pair of entities in the same row\\
     \citet{Venetis:2011:RST} & IS-A database & Frequency-based & Core + attribute columns \\
     \citet{Zhang:2017:EAE} & Freebase & Optimization & Any pair of columns \\
%      \citet{Zwicklbauer:2013:TDW} & DBpedia & Frequency-based & \\
    \bottomrule
  \end{tabular}
%  }
  \label{tbl:re:tasks}
\end{table}
Relation extraction refers to the task of associating a pair of columns in a table with the relation that holds between their contents and/or extracting relationship information from tabular data and representing them in a new format (e.g., RDF).
Table~\ref{tbl:re:tasks} summarizes the methods we will discuss below. 

\citet{Venetis:2011:RST} add annotations to tables to describe the binary relationships represented by columns.  This is accomplished by leveraging a relations database of (argument1, predicate, argument2) triples. For binary relationships, the relationship between columns $A$ and $B$ is labeled with $R$ if a substantial number of pairs of values from A and B occur in the relations database. \citet{Venetis:2011:RST} are only able to annotate a small portion of a whole table corpus (i.e., low recall). They discover that the vast majority of these tables are either not useful for answering entity-attribute queries, or can be labeled using a handful of domain-specific methods.

%We detail the column type identification task and table cell entity linking in Sect.~\ref{sub:sub:cti} and Sect.~\ref{sub:sub:el}, which involves table cell entity linking and column class label prediction. 
%By utilizing the results of entity linking and column type prediction, 
\citet{Mulwad:2010:ULD} propose a preliminary method for relation extraction, which utilizes the results of entity linking and column type prediction. Specifically, the method generates a set of candidate relations by querying DBpedia using SPARQL. Each pair of strings in two columns vote for the candidate relation. The normalized scores are used for ranking candidate relations and the highest one is taken as the column relation.
In follow-up work, \citet{Mulwad:2013:SMP} implement an improved semantic message passing method to extract RDF triples from tables. The semantic message passing first pre-processes the input table, separated by table elements such as column headers, cell values, columns, etc. Then, the processed table is passed to a \emph{query and rank} module, which turns to knowledge bases from Linked Open Data to find candidates for each table element. Finally, a \emph{joint inference} step uses a probabilistic graph model to rank candidate relations that were identified for the table elements. 
\citet{Mulwad:2013:SMP} point out that current methods rely on semantically poor and noisy knowledge bases and can only interpret part of a table (low recall). Moreover, systems for numeric values remain challenging, which is consistent with~\citep{Ibrahim:2016:MSE}.

TableMiner+~\citep{Zhang:2017:EAE} interprets relations between the core column and other columns on each row independently. It computes an individual confidence score for each candidate relation from each row. The candidate set of relations for two columns is derived by collecting the winning relations on all rows. A final confidence score of a candidate relation adds up its instance and context score computed based on context overlap. It is used to find the relation with the highest confidence.
A key finding in \citep{Zhang:2017:EAE} is that a system that is based on partial tabular data can be as good as systems that use the entire table.

Relation extraction can also be used to augment Linked Data repositories~\citep{Yoones:2014:KBA}. \citet{Yoones:2014:KBA} propose a probabilistic approach using under-explored tabular data. Assuming that the entities co-occurring in the same table are related, they focus on extracting relations between pairs of entities appearing in the same row of a table. Entities in table cells are mapped to a knowledge base first. Then, sentences containing both entities from the same table row are collected from a text corpus.  Next, textual patterns (describing the relationship between these two entities) are extracted.  Finally, the probability of the possible relations is estimated using Bayesian inference. A new relation, which is a triple consisting of two entities and a pattern, can be added to the Linked Data repository for augmentation. 
\citet{Munoz:2014:ULD} utilize entity annotations in Wikipedia tables.  Taking existing relations between entities in DBpedia, they look these entities up in Wikipedia tables.  This then indicates that the same relation stands between entities in other rows of this table.

%\citet{Zwicklbauer:2013:TDW} propose a simple method to annotate table headings with semantic types, using DBpedia's type system.  The method is divided into three steps: (i) table cell entity linking, using a search-based disambiguation method (detailed in Sect.~\ref{sec:interpret:el}), (ii) entity type resolution (looking up the corresponding entity types from DBpedia), and (iii) type aggregation, which takes the union of all entity types in that column and selects the most frequent of those as the type for the given table heading.
%Despite being considerably simpler, the method in \cite{Zwicklbauer:2013:TDW} achieves comparable accuracy to other methods, such as~\citet{Venetis:2011:RST}. The authors attribute it DBpedia being more exhaustive and containing high quality data.

%
\begin{figure}[t]
   \centering
   \includegraphics[width=0.8\textwidth]{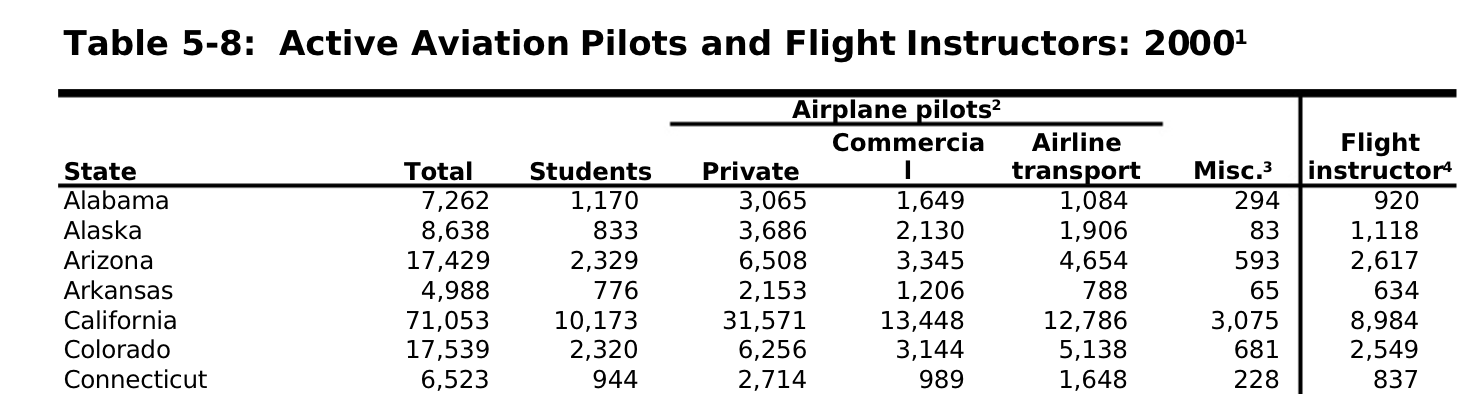} 
   \caption{Excerpt from a table containing hierarchical attributes. The example is taken from the U.S. Department of Transportation (\url{http://www.api.faa.gov/CivilAir/index.html}).}
\label{fig:spreadsheet}
\end{figure}

\citet{Chen:2013:AWS} introduce a system to automatically extract relational data from spreadsheets instead of the Web. 
Most of the methods on spreadsheets require users to provide sheet-specific rules~\citep{Ahmad:2003:TSS, Hung:2011:SCD}.  In contrast, \citet{Chen:2013:AWS} realize it in an automatic manner. 
Generally, the system detects attributes and values, identifies the hierarchical structure of attributes, and generates relational tuples from spreadsheet data. Specifically, the so-called \emph{frame finder} module of their system aims to identify the data frame regions within a spreadsheet.  These data frames consist of attribute and value regions.
First, it labels each row with one of the categories: title, header, data, or footnote. 
Then, the data frame regions are created, which are passed to the \emph{hierarchy extractor} for recovering the attribute hierarchies by finding all parent-child pairs in an attribute region. 
See Fig.~\ref{fig:spreadsheet} for an illustration, where \emph{Airplane pilots} and \emph{Airline transport} would be annotated as a parent-child attribute pair.
Finally, a series of parent-child candidates are generated and the true parent-child pairs are identified through classification. Alternatively, a so-called enforced-tree classification is proposed, which constructs a strict hierarchical tree for attributes. In the end, relational tuples are generated from the value region, whose value is annotated with one attribute from the attribute hierarchy.

\subsection{Other Tasks}

Data translation is concerned with the problem of mapping raw data, collected from heterogenous sources, to a transformed version for the end user~\cite{He:2018:TES}.
Tables encode a large number of mapping relationships as column pairs, e.g., person and birthday, which can be useful data assets for data translation.  \citet{Wang:2017:SMR} propose to automatically synthesize mapping relationships using table corpora by leveraging the compatibility of tables based on co-occurrence statistics. 
\citet{Braunschweig:2015:FWT} propose a method to normalize web tables in cases where multiple core columns and mixed concepts are detected in one table.

Web tables are embedded in HTML pages, where the surrounding text can help to understand what a given table is about.  However, these surrounding sentences are not equally beneficial for table understanding.  \citet{Wang:2015:CRW} present the Table-Related Context Retrieval system (TRCR) to determine the relevance between a table and each surrounding sentence.  Using TRCR, the most relevant texts are selected to uncover table semantics.  Another related study is performed in \cite{Govindaraju:2013:UTC}, where NLP tools, like part-of-speech tagging, dependency paths, and named-entity recognition, are explored to mine surrounding texts for understanding table semantics. 
\citet{Braunschweig:2015:CCE} propose a heuristic approach that extracts text snippets from the context of a web table, i.e., caption, headline, surrounding text, and full text, which describe individual columns in the table and link these new labels to columns. As a follow-up, \citet{Braunschweig:2016:PWT} propose a contextualization method of splitting table context into paragraphs with consistent topics, providing a similarity measure that is able to match each paragraph to the table in question. Paragraphs are then ranked based on their relevance.

% Head 1
%\section{Tasks}
%In this section, we \sz{list various} table related tasks. \todo{task-input-output-tabletype-ref}
% Head 2
%
\section{Table Search}
\label{sec:tablesearch}

Table search is the task of returning a ranked list of tables in response to a query. It is an important task on its own and is regarded as a fundamental step in many other table mining and extraction tasks as well, like table integration or data completion. 
Table search functionality is also available in commercial products; e.g., Microsoft Power Query provides smart assistance features based on table search.
Depending on the type of the query, table search may be classified as \emph{keyword-based search} and \emph{table-based search}. 

\subsection{Keyword-based Search} 
\label{sub:kqs}

Given a keyword query, the process of returning a ranked list of tables is referred to as \emph{keyword-based search} (or \emph{keyword query search}).
%According to the methodology, keyword query search is generally divided into \emph{using web search engine} and \emph{searching out of a table corpus}. 
%Using web search engine is to filter tables from returned web pages.
One of the first published methods is by \citet{Cafarella:2008:WEP}, who implement keyword table search on top of an existing web search engine.  Specifically, they extract the top-$k$ tables from the returned webpages.
In follow-up work, a similar system called OCTOPUS~\citep{Cafarella:2009:DIR} extends the same method (referred to as SimpleRank) with a reranking mechanism (SCPRank) that considers attribute co-occurrences.
%Instead of ending up with the top $k$ pages, \citet{Cafarella:2008:WEP} implements a filterRank method to go as far as possible to ensure to find top k tables.

Later works search directly within a designated table corpus.  Methods may be divided into \emph{document-based}  and \emph{feature-based} approaches.
According to the first group of approaches, a document-based representation is created for each table. This might  contain all text included in the table or only certain elements of the table (e.g., caption or header labels).  Then, these document-based representations may be ranked using traditional retrieval models, such as TF-IDF~\citep{Pimplikar:2012:ATQ}.

%\todo{[move this under table understanding]} To make table semantic machine processable, \citet{Wang:2012:UTW} also develop a table search engine for keyword query with the help of web table semantic understanding (detailed in Sec.~\ref{sub:sub:re}).\todo{[/move]}

\begin{table*}[t]
\centering
\caption{A selection of features for keyword-based table search.}
\begin{tabular}{p{2.8cm}p{8cm}p{4cm}}
	\toprule
	\multicolumn{2}{l}{\textbf{Query features}} & \textbf{Source}  \\
	\midrule
	QLEN & Number of query terms & \cite{Tyree:2011:PBR}  \\
	$\mathrm{IDF}_f$ & Sum of query IDF scores in field $f$ & \cite{Qin:2010:LBC}  \\	
	\midrule
	\multicolumn{3}{l}{\textbf{Table features}} \\
	\midrule
	\#rows & Number of rows in the table & \cite{Cafarella:2008:WEP,Bhagavatula:2013:MEM}\\
	\#cols & Number of columns in the table & \cite{Cafarella:2008:WEP,Bhagavatula:2013:MEM}  \\
	\#of NULLs in table & Number of empty table cells & \cite{Cafarella:2008:WEP,Bhagavatula:2013:MEM}  \\
	PMI & ACSDb-based schema coherency score & \cite{Cafarella:2008:WEP} \\	
	inLinks & Number of in-links to the page embedding the table & \cite{Bhagavatula:2013:MEM}  \\
	outLinks & Number of out-links from the page embedding the table & \cite{Bhagavatula:2013:MEM} \\
	pageViews & Number of page views & \cite{Bhagavatula:2013:MEM} \\
	tableImportance & Inverse of number of tables on the page & \cite{Bhagavatula:2013:MEM} \\
	tablePageFraction & Ratio of table size to page size & \cite{Bhagavatula:2013:MEM} \\
	%sectionNumber\cite{Bhagavatula:2013:MEM}&Wikipedia Section index the table occurs in&$[1,n]$\\
	\midrule
	\multicolumn{3}{l}{\textbf{Query-table features}} \\
	\midrule
	\#hitsLC & Total query term frequency in the leftmost column cells & \cite{Cafarella:2008:WEP}  \\
	\#hitsSLC & Total query term frequency in second-to-leftmost column cells & \cite{Cafarella:2008:WEP}  \\
	\#hitsB & Total query term frequency in the table body & \cite{Cafarella:2008:WEP}  \\
	qInPgTitle & Ratio of the number of query tokens found in page title to total number of tokens & \cite{Bhagavatula:2013:MEM}  \\
	qInTableTitle & Ratio of the number of query tokens found in table title to total number of tokens & \cite{Bhagavatula:2013:MEM}  \\
	yRank & Rank of the table's Wikipedia page in web search engine results for the query & \cite{Bhagavatula:2013:MEM}  \\
	MLM similarity & Language modeling score between query and multi-field document repr. of the table & \cite{Hasibi:2017:NTE}  \\
		\bottomrule
\end{tabular}
\label{tbl:features}
\end{table*}

Feature-based methods employ supervised machine learning for table ranking. 
Features may be divided into three main categories: query features, table features and query-table features. 
\emph{Query features} include query length and IDF scores of query terms. 
\emph{Table features} characterize the table in terms of its dimensions (number of rows, columns) and schema coherency.  With a focus on Wikipedia tables, \citet{Bhagavatula:2013:MEM} introduce features related to the  connectivity of the Wikipedia page (pageViews, inLinks, and outLinks) and the table's importance within the page (table importance and table page fraction).
Finally, \emph{query-table features} capture the degree of matching between the user's information need and the table.  Typically, these include similarity scores between the query and various table elements. Table~\ref{tbl:features} lists a selection of features for keyword table search.
In terms of learning algorithm, \citet{Cafarella:2008:WEP} train a linear regression classifier, while \citet{Bhagavatula:2013:MEM} train a linear ranking model learned with Coordinate Ascent.

% To easily compute the coherency score, the \emph{attribute statistics database} (AcsDB) \footnote{https://web.eecs.umich.edu/~michjc/data/acsdb.html} is made public that records corpus-wide statistics on co-occurrences of schema elements. 

Instead of relying on a single keyword query as input, \citet{Pimplikar:2012:ATQ} take $q$ columns, each described by a set of keywords $Q_1,\dots,Q_q$, as input (e.g., $Q_1=$``chemical element,'' $Q_2=$``atomic number,'' and $Q_3=$``atomic weight''), and return a table with $q$ columns as the answer.
First, they rank tables using the union of words in $Q_1,\dots,Q_q$.  Then, each table column is labeled with the query column it maps to. Finally, relevant columns and rows are merged into a single table, by considering the table-level relevance scores and the column-level mapping confidence scores. To decide if two rows are duplicates of each other, they employ the method in~\citep{Gupta:2009:ATA}.
\citet{Zhang:2018:AHT} perform semantic matching between queries and tables for keyword table search.  Specifically, they (1) represent queries and tables in multiple semantic spaces (both discrete sparse and continuous dense vector representations) and (2) introduce various similarity measures for matching those semantic representations.  For the former, both queries and tables are represented as bag-of-entities, bag-of-categories, word embeddings (trained on Google news) and graph embeddings respectively.  %Query entities are obtained by searching top-k related entities utilizing typical entity retrieval method. Table entities are from both the page title and the core column. Categories of table entities are used to generate the bag-of-categories. Entities are further replaced with graph embeddings and words in query and table are substituted by word embeddings to get graph embeddings and word embeddings.  
As for the latter, matching methods, they employ the early and late fusion patterns~\citep{Zhang:2017:DPF}.  They consider all possible combinations of semantic representations and similarity measures and use these as features in a supervised learning model.  They demonstrate significant and substantial improvements over a state-of-the-art feature-based baseline.
Most recently, \citet{Li:2019:TNW} train word embeddings utilizing the Wikipedia table corpus and achieve comparable results.

\subsection{Table-based Search}
\label{sub:tqs}
 
Table search is not limited to keyword queries.  The input may be also be a table, in which case the task of returning related tables is referred to as \emph{table-based search} (or \emph{query by table}).
At its core, this task boils down to computing a similarity score between the input and candidate tables, which we shall refer to as \emph{table matching}.  
Search by table may be performed for different goals: (1)  to be presented to the user to answer her information need~\citep{DasSarma:2012:FRT,Limaye:2010:ASW,Nguyen:2015:RSS} and (2) to serve as an intermediate step that feeds into other tasks, like table augmentation~\citep{AhmadovTELW:2015:THI,Lehmberg:2015:MSJ,Yakout:2012:IEA,Nargesian:2018:TUS}.

One group of approaches addresses the table matching task by using certain table elements as a keyword query, and scoring tables using keyword-based methods.  
For example, \citet{AhmadovTELW:2015:THI} use table entities and table headings as queries to retrieve a ranked list of tables for data completion (to be detailed in Sect.~\ref{sec:augment:dc}).  The two ranked lists are then intersected afterwards in order to arrive at a more complete result set.

\begin{table}[t]
  \centering
  \caption{Overview of table elements used when querying by table for various table-related applications. }
  \begin{tabular}{llllllll}
    \toprule
    \textbf{Reference}    & \textbf{Application}  & \textbf{$T_E$}     & \textbf{$T_H$}  & \textbf{$T_{[:,j]}$} & \textbf{$T_p$} & \textbf{$T_{[i,j]}$}\\
    \midrule
    \citet{AhmadovTELW:2015:THI} & Data completion   & \checkmark & \checkmark &         &         &         \\
    \citet{DasSarma:2012:FRT} & Schema complement       & \checkmark & \checkmark &         &         &         \\
     & Entity complement       & \checkmark &         &         &         &         \\
    \citet{Lehmberg:2015:MSJ} & Relation join      &         & \checkmark &         &         &         \\
    \citet{Limaye:2010:ASW} & Table cell retrieval         &         &         & \checkmark &         & \checkmark \\
    \citet{Nargesian:2018:TUS} & Table union search & \checkmark && \checkmark && \checkmark \\
    \citet{Nguyen:2015:RSS} & Diverse table search         &         & \checkmark &         &         & \checkmark \\
    \citet{Yakout:2012:IEA} & Table augmentation         &         & \checkmark & \checkmark & \checkmark & \checkmark \\
    \citet{Zhang:2019:RRT} & Table recommendation & \checkmark & \checkmark & \checkmark & \checkmark & \checkmark \\ 
    \bottomrule
  \end{tabular}
  \label{tbl:tablesearch:elements}
\end{table}

More commonly, table matching is tackled by dividing tables into various elements (such as table caption, table entities, column headings, cell values), then computing element-level similarities.  
Table~\ref{tbl:tablesearch:elements} provides an overview of the table elements that have been utilized in past work.  It is worth pointing out that in most of these cases, table search is not the ultimate goal, it is only used as a component in a larger application.
%the ultimate (end-to-end) task is not table search; therefore, some of the developed approaches are heavily tailored towards some specific application.
The Mannheim Search Join Engine~\cite{Lehmberg:2015:MSJ} seeks to extend the input table with additional attributes.  It utilizes table headings by comparing the column headings between the input table and candidate tables.  Specifically, they first filter tables that share at least one column heading with the input table, using exact term matching. Then, the table matching score is computed by (1) building an edit distance similarity matrix between the input and candidate tables' column headings, and (2) calculating the Jaccard similarity of the two tables using the matrix's maximum weighted bipartite matching score.
Similar to the Mannheim Search Join Engine that is based on table headings, \citet{Nargesian:2018:TUS} search tables that are likely unifiable with the seed table, which is called \emph{attribute union ability}. \citet{Nargesian:2018:TUS} formalize three statistical models to estimate the likelihood that two attributes contain values that are in the same domain. The simplest case, named \emph{set domains}, uses the size of the intersection of values between two columns. The second case, called \emph{semantic domains}, measures the semantic similarity between the values by mapping the columns to classes, e.g., entities. For values that are expressed in natural language, the third case of \emph{natural language domains} measures semantics based on natural langue rather than on ontologies. They use word embeddings trained based on Wikipedia documents to define natural language domains and statistical tests between the vectors are used to evaluate the likelihood that two attributes are from the same domain.
\citet{DasSarma:2012:FRT} aim to find related tables for augmenting the input table with additional rows or columns, referred to as \emph{entity complement} and \emph{schema complement}, respectively. 
Entity complement considers the relatedness between entity sets of the input and candidate tables.  Relatedness between two entities is estimated by representing entities as weighed label sets (from a knowledge base or from a table corpus) and taking their dot product.  \citet{DasSarma:2012:FRT} propose multiple methods to aggregate pairwise entity relatedness scores for computing relatedness between two sets of entities.
Schema complement combines two element-wise similarities: table entities and column headings. 
The former considers the overlap between table entities. The latter estimates the benefits of adding a column from the candidate table to the input table by determining the consistency between the new column and the existing columns of the input table.
\citet{Yakout:2012:IEA} propose InfoGather, a holistic method for matching tables in order to support three core operations: augmentation by column headings, augmentation by example, and column heading discovery. 
They consider element-wise similarities, including table context, URL, tuples, column headings, column values, and table data, as well as cross-element similarity between table and context.  Similarity is measured using the vector product of TF/IDF-weighted term vectors.  Then, element-level similarity scores are combined as features in a machine learned model.  In follow-up work, InfoGather is extended as InfoGather+~\cite{Zhang:2013:ISM} to incorporate tables with numeric and time-varying attributes.
\citet{Zhang:2019:RRT} perform table matching by representing table elements in multiple semantic spaces, and then combining element-level similarities using a discriminative learning model.

\citet{Nguyen:2015:RSS} consider the diversity of the returned tables.  They focus on two table elements: column headings and table data.  The former is similar in spirit to the Mannheim Search Join Engine~\citep{Lehmberg:2015:MSJ}.  The latter works by measuring the similarity between table columns, which are represented as term frequency vectors.

Unlike the above methods, which consider tables as the unit of retrieval, \citet{Limaye:2010:ASW} return a ranked list of cells as result.  
They train a machine learning method for annotating (1) entities in tables cells, (2) columns with types, and (3) relations between columns.  Then, search is performed by issuing an automatically generated structured query.

\section{Question Answering on Tables}
\label{sec:qat}

Tables are a rich source of knowledge that can be utilized for answering natural language questions. 
This problem has been investigated in two main flavors: (1) where the table, which contains the answer to the input question, is given beforehand~\citep{PasupatL:2015:CSP}, and (2) where a collection of tables are to be considered~\cite{Sun:2016:TCS}.  The latter variant shares many similarities with traditional question answering (QA), while the former requires different techniques.
One of the main challenges of QA on tables, shared by both scenarios, is how to match the unstructured query with the (semi-)structured information in tables. 
Question answering on tables is also closely related to work on natural language interfaces to databases, where the idea is that users can issue natural language queries, instead of using formal structured query languages (like SQL), for accessing databases~\cite{Ion:1995:NLI, Li:2014:CIN, Li:2005:NIN, Popescu:2003:TTN}. \emph{Semantic parsing} is the task of parsing natural language queries into a formal representation.  Semantic parsing is often used in question answering, by generating logical expressions that are executable on knowledge bases~\citep{Berant:2013:SPF,Fader:2014:OQA}.  

\subsection{Using a Single Table}
\label{sec:qat:single}

We first discuss approaches that take a single table as input (sometimes referred to as \emph{knowledge base table}~\citep{Yin:2016:NEL}), and seek the answer to the input question in that table. 
The basic idea is to regard the input table as a knowledge base, which poses a number of challenges. 
First, knowledge bases contain a canonicalized set of relations, while tabular data is much more noisy. 
Second, traditional semantic parsing sequentially parses natural language queries into logical forms and executes them against a knowledge base.  To make them executable on tables, special logical forms are required.
Lastly, semantic parsing and query execution become complicated for complex questions as carefully designed rules are needed to be able to parse them into logical forms.
\citet{PasupatL:2015:CSP} propose to answer complex questions, involving operation such as comparison, superlatives, aggregation, and arithmetics in order to address above problems. They convert the input table into a knowledge graph by taking table rows as row nodes, strings as entity nodes, and columns as directed edges. The column headings are used as predicates. Numbers and strings are normalized following a set of manual rules.  Being one of the earliest works addressing this task, \citet{PasupatL:2015:CSP} follow a traditional parser design strategy.  A semantic parser is trained based on a set of question-answer pairs.  Given a table and a question, a set of candidate logical forms is generated by parsing the question. Then, logical forms are ranked using a feature-based representation, and the highest ranked one is applied on the knowledge graph to obtain the answer.
\citet{PasupatL:2015:CSP} develop a dataset, called WikiTableQuestion, which consists of a random sample of 2,100 tables from Wikipedia and 22,000 question-answer pairs. 
The proposed approach is found to suffer from a coverage issue, i.e., it is able to answer only 20\% of the queries that have answers in Freebase.

Different from semantic parsing methods that require predefined logical operations, 
\citet{Yin:2016:NEL} propose a neural network architecture, named Neural Enquirer, for semantic parsing with a specific table.  Neural Enquirer is a fully neural system that generates distributed representations of queries and tables (called query encoder and table encoder, respectively). Then, the question is executed through a series of query operations, called executors, with intermediate execution results computed in the form of table annotations at different levels.  Training can be performed in an end-to-end fashion or carried out using step-by-step supervision (for complex queries).
They use query answers as indirect supervision, but jointly perform semantic parsing and query execution in distributional spaces. The distributed representations of logical forms are learned in end-to-end tasks, which is based on the idea of adopting the results of the query execution as indirect supervision to train the parser. 
It is worth pointing out that this model makes a number of strong assumptions. For example, they consider four specific types of queries, provide the logical form template for each, and  carefully and manually select a table that is supplied as part of the input.
Similar to \citep{Yin:2016:NEL}, 
\citet{Arvind:2015:NPI} attempt to solve the task of question answering on tables using neural networks, a system called Neural Programmer.  Neural Programmer runs for $T$ steps and the final result is formed step by step.  The model adopts a Recurrent Neural Network (RNN) architecture to process the input question, a selector to assign probabilities to a set of possible operations and data segments at each step, and a history RNN to remember the previous operation and data selections. Providing a small set of basic operations, they also take the result of correct executions as indirect supervision. During the training, adding random noise to the gradient greatly improves performance as the operations and data selections are quite heterogenous.

\subsection{Using a Collection of Tables}

Another line of work focuses on answering questions using a collection of tables.
These approaches are more similar to traditional question answering on text, comprising of candidate identification, query type prediction, and ranking components.  The main differences are twofold. One is schema matching, which is the same as before (in Sect.~\ref{sec:qat:single}), but there is an additional normalization issue across tables here. The other is the need for extracting quantity values from tables.
\citet{Sarawagi:2014:OQQ} show that over 40\% of table columns contain numeric quantities, and propose a collective extraction framework to extract quantities from raw web tables based on a consensus model. 
Their system, called QEWT, extends the work of \citet{Banerjee:2009:LRQ} to tables. 
They employ keyword-based table ranking in order to fetch table candidates. This corresponds to the candidate snippet/answer identification step in traditional QA.
QEWT can answer quantity-target queries with a ranked list of quantity distributions, which are taken from the tables.
It uses a table column unit annotator based on probabilistic context free grammars for easily extracting quantities from table columns to deal with ambiguity (both for headings and for values). From an information retrieval perspective, quantity queries on web tables is the task of returning a ranked list of quantities for a query. QEWT employs a quantity response model for this task. 

Inspired by classical textual QA,
\citet{Sun:2016:TCS} decompose table cells into relational chains, where each relational chain is a two-node graph connecting two entities. 
Specifically, each row of a table represents relations among the cells. They construct a pseudo-node to connect all the cells and take the headings to label relationships. Any pair of cells in the same row form a directional relational chain.
The input query is also represented as a two-node graph question chain, by identifying the entities using an entity linking method. The task then boils down to finding the relational chains that best match the question chain.  This matching is performed using deep neural networks, in particular the Convolutional Deep Structured Semantic Model (C-DSSM)~\citep{Shen:2014:LSR}, to overcome the vocabulary gap limitation of bag-of-words models. 
They find that combining the deep features with some shallow features, like term-level similarity between query and table chains, achieve the best performance.
\citet{Sun:2016:TCS} conclude that their method can complement KB-based QA methods by improving their coverage.

\section{Knowledge Base Augmentation}
\label{sec:kba}

The knowledge extracted from tabular data can be used for enriching knowledge bases.  First, we present an approach that is devised for exploring the knowledge contained in web tables.  Then, we discuss methods for knowledge base augmentation using tabular data.

\subsection{Tables for Knowledge Exploration}
\label{sub:tm:tke}

\begin{figure*}[t]
	\centering
	\begin{tabular}{cc}
		\subfigure[]{\includegraphics[width=0.45\textwidth]{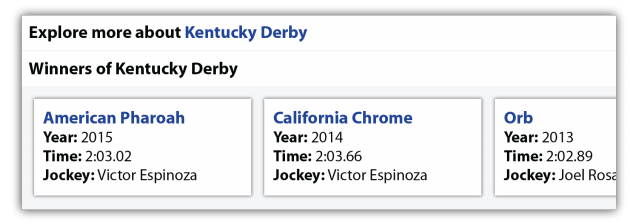}\label{fig:ke_a}}
		&
		\subfigure[]{\includegraphics[width=0.45\textwidth]{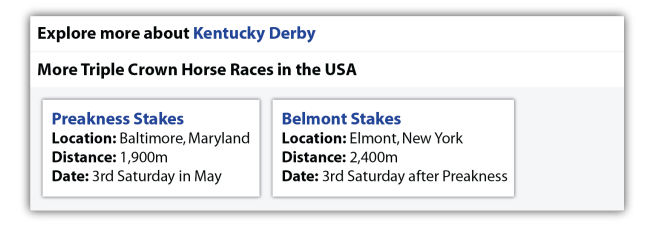}\label{fig:ke_b}}
	\end{tabular}
%	\vspace*{-\baselineskip}
	\caption{Illustration in~\citep{Chirigati:2016:KEU}, showing an example of knowledge exploration for the query of ``kentucky derby'' through Knowledge Carousels: (a) a downward showing the winners of Kentucky Derby; (b) a sideway representing the famous Triple Crown horse races in the US, of which Kentucky Derby is a member.}
	 \label{fig:ke}
\end{figure*}

The knowledge contained in web tables can be harnessed for knowledge exploration.
Knowledge Carousels~\cite{Chirigati:2016:KEU} is the first system addressing this, by providing support for exploring ``is-A'' and ``has-A'' relationships. These correspond to two kinds of entity-seeking queries (queries searching for attributes and relationships of entities), called ``sideways'' and ``downwards,'' respectively. Given an input entity, \citet{Chirigati:2016:KEU} utilize web tables to select the carousel type, create a set entities of this carousel, generate human-readable titles, and rank carousels based on popularity and relatedness extracted from tables.  See Fig.~\ref{fig:ke} for an illustration.

\subsection{Knowledge Base Augmentation and Construction}
\label{sub:tm:kba}
Tabular data on the Web can be used to construct new or augment existing knowledge bases.

\subsubsection{Knowledge Base Augmentation}
\label{sub:kba:kbma}
%The content and topical areas of tables are quite various. Much attention is paid to matching web tables against existing knowledge bases for better understanding them, which in return 
In Sect.~\ref{sec:interpret}, we have presented techniques for interpreting tables with the help of knowledge bases. The obtained annotations, in turn, can 
contribute to extending those knowledge bases. \emph{Knowledge base augmentation}, also known as \emph{knowledge base extension}, is concerned with generating new instances of relations using tabular data and updating knowledge bases with the extracted information.

Knowledge bases need to be complete, correct, and up-to-date. A precondition of extending knowledge bases using web tables is matching them to those existing knowledge bases. 
Specific matching problems include \emph{table-to-class matching}, \emph{row-to-instance matching}, and \emph{attribute-to-property matching}.
\citet{Ritze:2015:MHT} propose an iterative matching method, T2K, to match web tables to DBpedia for augmenting knowledge bases.  They also develop and make publicly available the T2D dataset for matching, consisting of 8,700 schema-level and 26,100 entity-level correspondences between web tables and DBpedia, which are extracted and annotated manually.  The T2K method utilizes the T2D dataset to execute iterative steps between candidate matching and property matching, to find proper entities/schemas in DBpedia for table rows/columns. 
However, T2D mainly focuses on large tables and does not work that well for small-sized tables~\citep{Lehmberg:2017:SWT}. To counter this problem, \citet{Lehmberg:2017:SWT} propose to combine tables from each website into larger tables for table matching, building on the intuition that tables from the same website are created in a similar fashion.

Strictly speaking, we classify the work in~\citep{Wang:2015:CEU} as \emph{row extension}.  Nevertheless, since they map table entities to a knowledge base with the purpose of collecting more entities from other tables that belong to the same concept in the knowledge base, their work can also be regarded as a knowledge base augmentation approach.

\subsubsection{Knowledge Base Construction}
Instead of augmenting existing knowledge bases, web tables contain abundant information to be turned into knowledge bases themselves. % or can be exploited to construct individual knowledge bases.

Even though there exists a number of large-scale knowledge bases, they are still far from complete~\citep{Dong:2014:KVW}. Therefore, \citet{Dong:2014:KVW} introduce a web-scale probabilistic knowledge base named Knowledge Vault (KV) that fuses different information sources. For web tables, \citet{Dong:2014:KVW} firstly identify the relation that is expressed in a table column by checking the column's entities, and reason about which predicate each column could correspond to.  This latter task is approached using a standard schema matching method~\citep{Venetis:2011:RST}, with Freebase as the knowledge base. The extracted relations, together with relational data from other sources, are converted into RDF triples, along with associated confidence scores. The confidence scores are computed based on a graph-based method. Specifically, the triples are fused by machine learning methods from multiple sources, including an existing knowledge base, (i.e., Freebase) and web tables. Consequently, 1.6B triples are generated, of which 324M have a confidence score above 0.7 and 271M have a confidence score above 0.9. 
%\citet{Dong:2014:KVW} compare different extraction methods and prior probabilistic methods as one contribution in this study as well.

\section{Table Augmentation} %/Table Extension}
\label{sec:augment}

\begin{figure}[!tbp]
   \centering
   \includegraphics[width=0.75\textwidth]{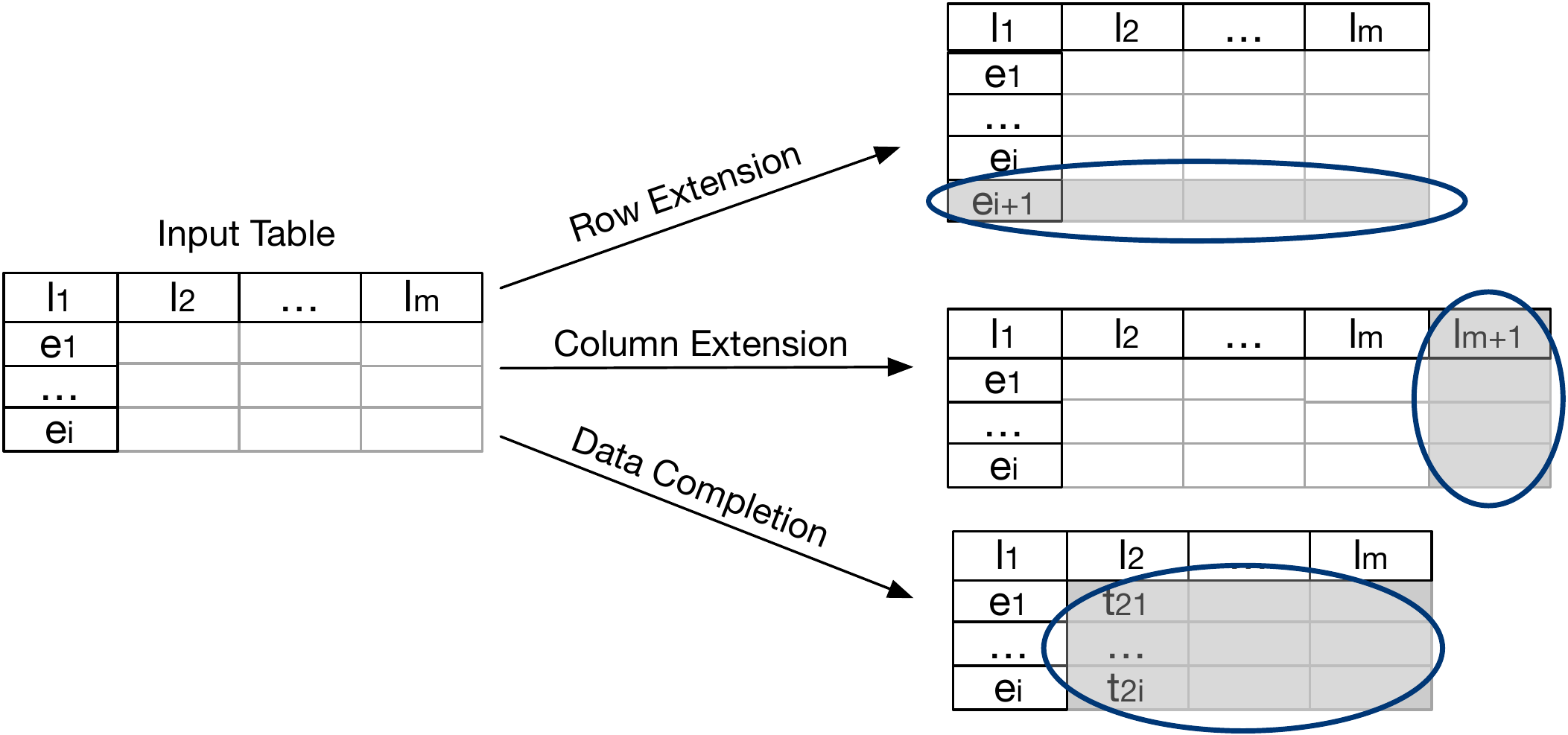} 
   \caption{Illustration of three table augmentation tasks: row extension, column extension, and data completion.} % \todo{Subdivide Column extension into (a) only heading label (b) heading + values; Subdivide data completion into (a) join and (b) data imputation}
\label{fig:table_augmentation}
\end{figure}

\emph{Table augmentation} refers to the task of extending a seed table with more data. 
Specifically, we discuss three tasks in this section: row extension (Sect.~\ref{sec:augment:re}), column extension (Sect.~\ref{sec:augment:ce}), and data completion (Sect.~\ref{sec:augment:dc}). See Fig.~\ref{fig:table_augmentation} for an illustration.
One might envisage these functionalities being offered by an intelligent agent that aims to provide assistance for people working with tables~\citep{Zhang:2017:ESA}.

\subsection{Row Extension}
\label{sec:augment:re}

\begin{table}[t]
  \centering
  \caption{Overview of row population methods. Notice that table search is inherently involved.}
%  {\color{blue}
  \begin{tabular}{lcccccc}
    \toprule
    & \multicolumn{2}{c}{\textbf{Data}} &&& \multicolumn{2}{c}{\textbf{Tasks}} \\
    \cline{2-3} \cline{6-7}
    \textbf{Reference} & \textbf{KB} & \textbf{Tables} &&& \textbf{Table search} & \textbf{Row population} \\
    \midrule
       \citet{DasSarma:2012:FRT} & & \checkmark &&& \checkmark \\
       \citet{Wang:2015:CEU} & & \checkmark &&& \checkmark & \checkmark*      \\
       \citet{Yakout:2012:IEA} & & \checkmark &&& \checkmark & \checkmark \\
       \citet{Zhang:2017:ESA} & \checkmark & \checkmark &&& \checkmark & \checkmark \\
    \bottomrule
    \multicolumn{7}{l}{$^*$ \footnotesize{Originally developed for concept expansion, but can be used for row population.}}
  \end{tabular}   
  \label{tbl:tableaug:row}
%  }
\end{table}
\begin{figure}[!tbp]
   \centering
   \includegraphics[width=0.75\textwidth]{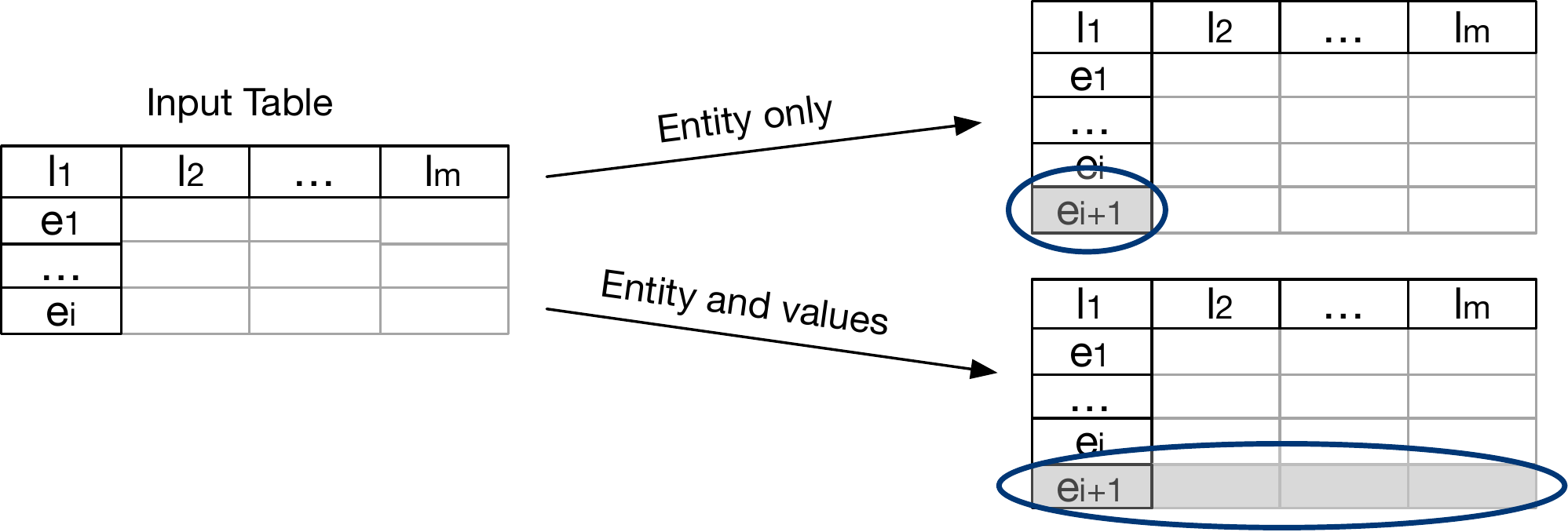} 
   \caption{Illustration of row extension by adding only an entity (Upper) or an entire row, i.e., an entity as well as cell values (Lower). Existing work has focused on the former task.}
\label{fig:row-extension}
\end{figure}

\emph{Row extension} aims to extend a given table with more rows or row elements (see Fig.~\ref{fig:row-extension}). It mainly focuses on a particular type of tables, namely, relational tables.  More specifically, row extension primarily targets horizontal relational tables, where rows represent entities and columns describe the attributes of those entities.  In such tables there usually exists a \emph{core column} (or \emph{key column}) containing mostly entities~\citep{Bhagavatula:2015:TEL,Venetis:2011:RST}.  Instead of directly providing a complete tuple (row), existing work has focused on identifying entities for populating such core columns (i.e., the Upper scenario in Fig.~\ref{fig:row-extension}).
Table~\ref{tbl:tableaug:row} provides an overview of methods that will be covered below. As we shall see, table search is inherently involved here.

Populating entities in the core column of a table is similar to the problem of \emph{concept expansion}, also known as \emph{entity set expansion}, where a given set of seed entities is to be completed with additional entities~\cite{Bron:2013:EBE, He:2011:SEI,Metzger:2013:QQE,Metzger:2014:ASE}.
Existing methods for concept expansion suffer from two main issues: input ambiguity and semantic drift (i.e., entities belonging to different concepts are mixed during expansion).
Motivated by the intuition that tables tend to group entities that belong to a coherent concept, \citet{Wang:2015:CEU} leverage web tables for the concept expansion task, thereby aiming to prevent semantic drift.  
They provide both the seed entities as well as a concept name as input.  First, they retrieve tables related to the seed entities. 
%\sz{\citet{Wang:2015:CEU} find that the existing holistic ranking methods like PageRank will result in semantic drift. Therefore,} 
Then, they use a graph-based ranking method to rank candidate entities that co-occur with the seed entities in those tables. Specifically, they first expand the set by iteratively adding the most relevant tables based on concept likelihood, and collecting entities there. Then, they refine the earlier estimation and remove less relevant tables  based on more complete information. \citet{Wang:2015:CEU} find that adding an input concept can address the semantic drift problem for tail concepts.
While this method is developed for concept expansion, it is directly applicable to the problem of populating entities in a core column.

% yet the input might be slight different. A plenty of work implementing entity set expansion utilising the external web resources. However, we just detail table-related tasks here, where table elements are the input.

\citet{DasSarma:2012:FRT} search for \emph{entity complement} tables that are semantically related to entities in the input table (as we have already discussed in Sect.~\ref{sub:tqs}).  Then, the top-$k$ related tables are used for populating the input table. \citet{DasSarma:2012:FRT}, however, stop at the table search task.
A similar approach is taken in InfoGather~\citep{Yakout:2012:IEA}, where this task is referred to as the \emph{augmentation by example} operation.  There, they first search for related tables (cf. Sect.~\ref{sub:tqs}), and then consider entities from these tables, weighted by the table relatedness scores. \citet{Yakout:2012:IEA} build a schema matching graph among web tables, based on pairwise table similarity.  Despite the use of scalable techniques, this remains to be computationally very expensive, which is a main limitation of the approach.
Instead of relying only on related tables from a table corpus, \citet{Zhang:2017:ESA} also consider a knowledge base (DBpedia) for identifying candidate entities.  Specifically, they collect entities sharing the same types or categories with the input entities from DBpedia, and entities from similar tables (i.e., tables sharing seed entities, having similar captions, or including the same headings) as candidates. They find that entity type information in DBpedia is too general to help identify relevant candidates, and end up using only category information when extracting candidates from DBpedia.  
It is also shown that using related tables and using a knowledge base are complementary when identifying candidate entities.
They develop a generative probabilistic model for the subsequent ranking of candidate entities based on their similarity to (1) other entities in the table, (2) column headings, and (3) the caption of the input table. Among the three table elements, seed entities are the most important component for entity ranking, followed by table headings and caption. A combination of the three table elements performs the best in the end.
In recent work, ~\citet{Li:2019:TNW} utilize Word2vec to train table embeddings for core column entities. Combining the embedding-based similarity scores with the probability-based scores from~\citep{Zhang:2017:ESA} results in further performance improvements.

% \citet{DasSarma:2012:FRT} focus on tables that have a set of entities. Entity complement is a table relatedness method measuring entity sets between tables. The table with the best related table entity set will be the candidate and its entities are the candidates be to added to the seed table. In ~\cite{DasSarma:2012:FRT}, an entity set are presented as the label vectors and their similarity is computed above them.

% Is is more convenient if only ranked entities other than tables (e.g., \cite{DasSarma:2012:FRT}) are outputted for row extension. The most fit method to this task is taken by ~\cite{Zhang:2017:ESA}, where the row extension is defined as \emph{row population}. Comparing to the above method, ~\citet{Zhang:2017:ESA} directly provide a ranked list of entities to the seed table. In ~\cite{Zhang:2017:ESA}, the row population assistance is only offered to entity-focused tables, whose leftmost column contains only entities as values and those entities are unique within the column. \citet{Zhang:2017:ESA} filter the candidate entities from DBpedia and relevant tables at the first step and then develop generative probabilistic models considering entity, table column and table caption to rank these candidates. A more appropriate method is combining the above two methods to fetch the most fit value.

\subsection{Column Extension}
\label{sec:augment:ce}

\begin{figure}[!tbp]
   \centering
   \includegraphics[width=0.75\textwidth]{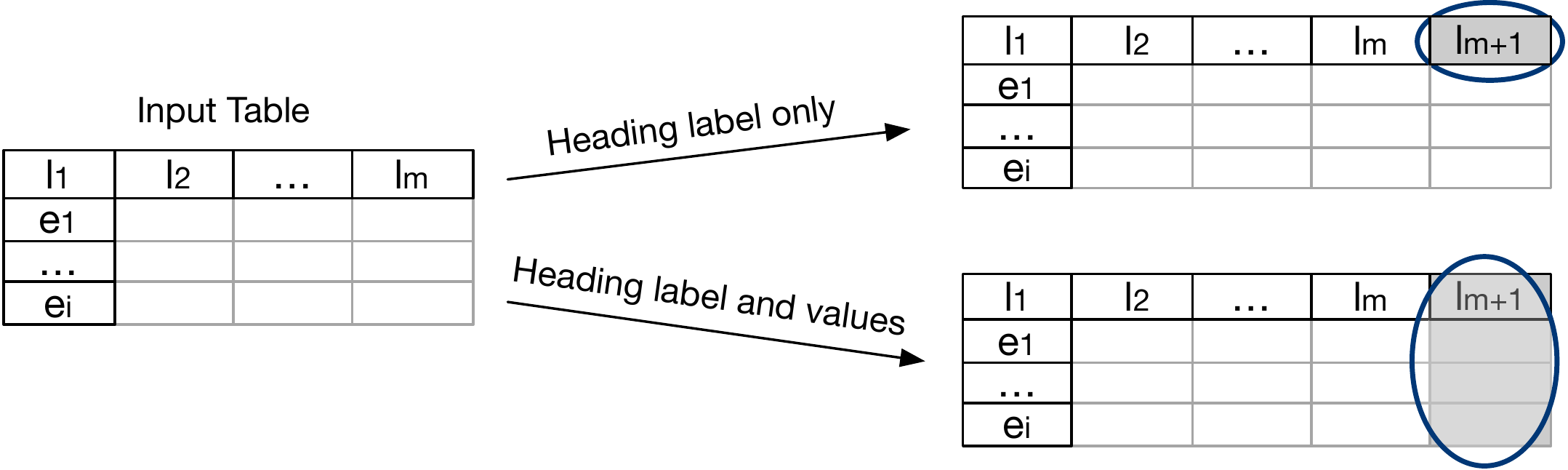} 
   \caption{Illustration of column extension by adding only a heading label  (Upper) or an entire column, i.e., heading label and cell values (Lower).}
\label{fig:column-extension}
\end{figure}
\begin{table}[t]
  \centering
  \caption{Overview of column population methods.}
%  \small
%  {\color{blue}
  \begin{tabular}{llccccccc}
    \toprule
    && \multicolumn{3}{c}{\textbf{Data}} &&& \multicolumn{2}{c}{\textbf{Output}} \\
    \cline{3-5} \cline{7-8}
    \textbf{Reference} & \textbf{Task} & \textbf{Web tables} & \textbf{WP tables} & \textbf{KBs} && \textbf{$T_H$} & \textbf{$T_H + T_{[:,j]}$}  \\
    \midrule
      \citet{Bhagavatula:2013:MEM} & Relevant join & &\checkmark&&& & \checkmark \\
      \citet{Cafarella:2008:WEP} & Schema auto-completion & \checkmark &&&&  \checkmark &   \\
      \citet{DasSarma:2012:FRT} & Schema complement & \checkmark&&&& \\
      \citet{Lehmberg:2015:MSJ}  & Search join & \checkmark&\checkmark&\checkmark&& & \checkmark \\
      \citet{Zhang:2017:ESA} & Column population & &\checkmark&&& \checkmark \\
    \bottomrule
  \end{tabular}
  \label{tbl:tableaug:col}
%  }
\end{table}

The most widely studied subtask in table augmentation is \emph{column extension}: extending a table with additional columns. 
This task roughly corresponds to the \emph{join} operation in databases.  In this context, the set of column heading labels is also often referred to as the table \emph{schema}.
Commonly, column extension is approached by first locating similar tables and then considering the column headings/values in those tables.
Table~\ref{tbl:tableaug:col} provides an overview of the methods discussed in this section.

% JUST LABELS
One particular variant of column extension aims to identify additional column heading labels (see Fig.~\ref{fig:column-extension} (Upper)).  As table columns often correspond to entity attributes, this task is also referred to as \emph{attribute discovery}~\citep{Yakout:2012:IEA} or \emph{schema auto-complete}~\citet{Cafarella:2008:WEP}.
The WebTables system~\cite{Cafarella:2008:WEP} implements this functionality based on the \emph{attribute correlation statistics database} (ACSDb).  ACSDb contains frequency statistics of attributes and co-occurring attribute pairs in a table corpus.  ACSDb comprises 5.4M unique attribute names and 2.6M unique schemas.
With these statistics at hand, the next probable attribute can be chosen using a greedy algorithm.
The statistics-based method in~\cite{Cafarella:2008:WEP} was the first approach to column extension, and was found to be able to provide coherent heading suggestions. However, later research has proven that considering additional features can further improve performance.
\citet{DasSarma:2012:FRT} focus on finding related tables, with the aim of schema complement. For ranking tables, they consider two factors: (1) the coverage of entities, and (2) the potential benefits of adding additional attributes from those tables (we discussed the table search method in Sect.~\ref{sub:tqs}).  Again, they stop at the table search task.
The task of identifying potential attributes or column labels is also known as \emph{schema matching}~\citep{Lehmberg:2015:MSJ} or \emph{column population}~\citep{Zhang:2017:ESA}.
\citet{Zhang:2017:ESA} try to find the headings that can be placed as the next column in an input table. They first find candidate headings from similar tables (the same strategy that they also use for row population). \citet{Zhang:2017:ESA} observe that input entities and table caption contribute comparably to the identification of relevant candidates, while table headings are the least important component. However, similar to row population, all these sources are complementary, i.e., each source can identify candidate headings that none of the others could.  In a subsequent ranking step, the candidates are scored based on table similarity, by aggregating element-wise similarities between (corresponding elements of) the input table and related tables.
In~\citep{Li:2019:TNW}, they utilize Word2vec to train embeddings for table headings. Similar to row population, combining the embedding similarity scores with the probabilities from~\citep{Zhang:2017:ESA} yields further performance improvements.
The above approaches differ in what they use as input, i.e., whether they use only table headings~\citep{Cafarella:2008:WEP,Lehmberg:2015:MSJ} or the entire table~\citep{DasSarma:2012:FRT,Zhang:2017:ESA}.

%\sz{Essentially,  the schema matching in~\cite{Lehmberg:2015:MSJ} is used to return the potential attributes or column labels instead of complete columns. With a focus on tables headings extension, \citet{Zhang:2017:ESA} re-define this task as \emph{column population}. Given a seed table, mainly taking the existing column labels as input, \emph{column population} returns a ranked list of lables to be added into the seed table from the information retrieval perspective. In~\cite{Zhang:2017:ESA}, the seed table is fully utilised, they develop a generative probabilistic model taking table caption, table entity set (leftmost column) and seed table headings as input to rank the candidate labels filtered from relevant tables from an early step.
%}

% ONLY VALUES => moved to data completion

% ENTIRE COLUMN, INCLUDING LABEL+VALUES

Another variant attempts to augment the input table with entire columns, that is, including both the heading label as well as the corresponding cell values for each row within that column (see Fig.~\ref{fig:column-extension} (Lower)).
\citet{Bhagavatula:2013:MEM} present the \emph{relevant join} task, which returns a ranked list of column triplets for a given input table. Each triplet consists of \emph{SourceColumn}, \emph{MatchedColumn}, and \emph{CandidateColumn}. \emph{SourceColumn} is from the query table, while \emph{MatchedColumn} and \emph{CandidateColumn} are from the candidate tables. 
They propose a semantic relatedness measure to find candidate tables from related Wikipedia pages, where page relatedness is estimated based on in-link intersections.
Their idea is to compute similarity between columns, such that if \emph{SourceColumn} and \emph{MatchedColumn} share largely similar values, then the input table may be extended with \emph{CandidateColumn}.   These candidate columns are classified as relevant or non-relevant, using a linear ranking model, before performing the actual join.  
%In~\cite{Bhagavatula:2013:MEM}, the semantic relatedness, which compares the \sz{Wikipedia} pages' relatedness, is adopted to enhance the performance.
To reduce the number of candidate columns, some are filtered out in a pre-processing stage using simple heuristics.  Columns that are kept are required to be non-numeric, have more than four rows, and an average string length larger than four. 
\citet{Bhagavatula:2013:MEM} find that columns containing numeric data make more relevant additions than non-numeric ones. Additionally, more distinct values in the \emph{SourceColumn} and a higher match percentage lead to better quality joins.
The join operation is also supported by the Mannheim Search Join Engine~\cite{Lehmberg:2015:MSJ}. It first searches for related tables based on column headings (cf. Sec.~\ref{sub:tqs}), then applies a series of left outer joins between the query table and the returned tables. Afterwards, a consolidation operation is performed to combine attributes. Specifically, they employ a matching operator that relies on data from knowledge bases. Given two columns, similar match (Levenshtein distance) and exact match are used for matching headings.
\citet{Lehmberg:2015:MSJ} observe that similar match returns on average 3.4 times more tables than exact match. Among different table corpora, web tables provide the largest number of relevant tables, and Wikipedia tables tend to be populous on certain topics, such as countries and films.

%A number of study are carried out from a data base perspective. Meanwhile, considering column as a search target, this task could be solved as an information retrieval problem. 

\subsection{Data Completion}
\label{sec:augment:dc}

\begin{figure}[t]
   \centering
   \includegraphics[width=0.75\textwidth]{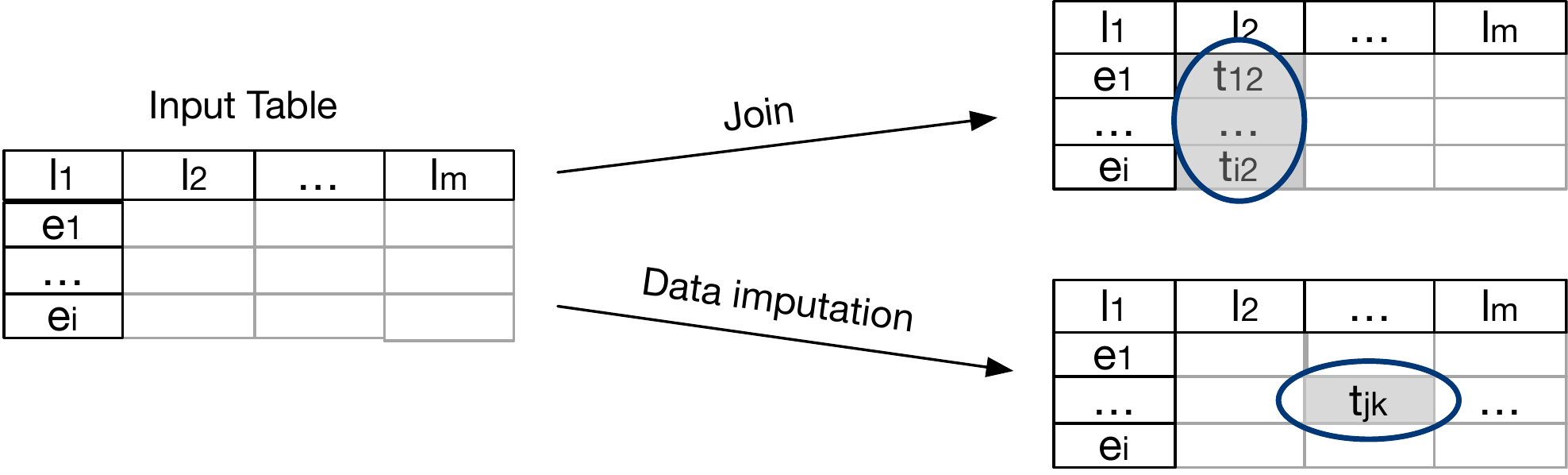} 
   \caption{Illustration of data completion tasks: join (Upper) and data imputation (Lower).}
\label{fig:data-completion}
\end{figure}
\begin{table}[t]
  \centering
  \caption{Overview of data completion methods.}
%  \small
%  {\color{blue}
  \begin{tabular}{lccccc}
    \toprule
    & \multicolumn{2}{c}{\textbf{Data}} && \multicolumn{2}{c}{\textbf{Output}} \\
    \cline{2-3} \cline{5-6}
	\textbf{Reference}  & \textbf{Tables} & \textbf{Web} && \textbf{$T_{[:,j]}$} & \textbf{$T_{[i,j]}$}  \\
    \midrule
      \citet{AhmadovTELW:2015:THI}  &  \checkmark & \checkmark && \checkmark & \checkmark  \\
      \citet{Cafarella:2009:DIR}  &  \checkmark & && \checkmark & \\
      \citet{Yakout:2012:IEA}  & \checkmark & && \checkmark & \\
      \citet{Zhang:2013:ISM}  &  \checkmark & && \checkmark & \\
      \citet{Zhang:2019:ADC}  & \checkmark & \checkmark & && \checkmark \\
    \bottomrule
  \end{tabular}
  \label{tbl:tableaug:data}
%  }
\end{table}

\emph{Data completion} for tables refers to the task of filling in the empty table cells. 
Table~\ref{tbl:tableaug:data} summarizes the methods we discuss here.
One variant of this task attempts to find the cell values for an entire column (see Fig.~\ref{fig:data-completion} (Upper)). This is known as the \emph{augmentation by attribute name} operation in the InfoGather system~\citep{Yakout:2012:IEA}. %; see Fig.~\todo{X (b)}.
This is typical of a scenario where the core entity column as well as the column headings are given in a relational table, and the values for the corresponding attributes (\emph{augmenting attributes}) are to be filled in. 
The system in \citep{Yakout:2012:IEA} takes the incomplete table as input to search for matching tables, then extracts attribute values from those tables.
%The two operations of InfoGather~\cite{Yakout:2012:IEA}, entity augmentation by example and augmentation by attribute name, though we classify them into row extension, when the missing table cells are entities, they can also be applied to data completeness task. 
It is worthwhile to point out that InfoGather focuses on finding values that are entities.
An extended version of the system, InfoGather+~\cite{Zhang:2013:ISM}, focuses on numerical and time-varying attributes.
They use undirected graphical models and build a semantic graph that labels columns with units, scales, and timestamps, and computes semantic matches between columns. Their experiments are conducted on three types of tables: company (revenue and profit), country (area and tax rate) and city (population). \citet{Zhang:2013:ISM} find that the conversion rules (manually designed unit conversion mapping) achieve higher coverage than string-based schema matching methods.   %InfoGather+ is potentially useful for string or entity values as well.}
%Therefore, the systems of InfoGather and InfoGather+ can be combined to find missing values for table when entity and numeric cells needed.
Similar to InfoGather's \emph{augmentation by attribute name} operation, the \emph{extend} operation in the OCTOPUS systems~\cite{Cafarella:2009:DIR} enables the user to add more columns to a table by performing a join. It takes a keyword query and a given (existing) table column as input, where the keyword describes the newly added column.  Different from a regular join, the added column is not necessarily an existing column.  It may be formed row-by-row by combining information from multiple related tables (see Sect.~\ref{sub:kqs} for the table search operation). However, \citet{Cafarella:2009:DIR} rely on simple methods like edit distance for schema matching, which leaves room for improvement.

Another flavor of the data completion task focuses on filling in missing values for individual cells, referred to as \emph{data imputation} (see Fig.~\ref{fig:data-completion} (Lower)).  
\citet{AhmadovTELW:2015:THI} present a hybrid imputation method that combines a lookup-based approach, based on a corpus of web tables, and a model-based approach that uses machine learning (e.g., k-nearest neighbors or linear regression) to predict the value for a missing cell. 
It is worth noting that all the above methods rely only on tables and ignore the cases where no similar tables can be found.
The method in~\citep{AhmadovTELW:2015:THI} is shown to be able to improve coverage.  However, being able to automatically decide when to do simple lookup and when to employ a machine learned model remains an open challenge.
CellAutoComplete is a recent framework proposed by~\citet{Zhang:2019:ADC} to tackle several novel aspects of this problem, including: (1) enabling a cell to have multiple, possibly conflicting values, (2) supplementing the predicted values with supporting evidence, (3) combining evidence from multiple sources, and (4) handling the case where a cell should be left empty. This framework makes use of a large table corpus and a knowledge base as data sources, and consists of preprocessing, candidate value finding, and value ranking components.
%\sz{The way to combine the Web and tables as \citet{AhmadovTELW:2015:THI} will definitely improve the coverage. 
% However, \citet{AhmadovTELW:2015:THI} fail to find out an automatic manner to select machine learning models as they propose.} 

% Possibly, some table elements from the matched table can be added to the input table.  However, we can not always find the well matched and existing tables. Web tables are storing abundant information, existing data instead of a whole table also contribute to this task.  For those tables cells out of looking up, \citet{AhmadovTELW:2015:THI} develop a data lake over Web tables and train a machine learning method to predict a value for the missing cell.

% Start of "Sample References" section

%\section{Other Tasks}
%\label{sec:other}

%In the field of data base, there are still various tasks focusing on web tables, as we mentioned at the beginning, we limit our focus to the perspective to information retrieval, and therefore many studies from other fields are not included in this study.

\section{Conclusions and Future Directions}
\label{sec:concl}

Tables are a powerful and popular tool for organizing and manipulating data.  Research on web tables has seen nearly two decades of development.  During this long period, the research focus has evolved considerably, from low level tasks (table extraction) to tapping more and more into the actual knowledge contained in tables (for search and for augmenting existing knowledge bases).  
Below, we review past progress and identify open research directions for each of the six main categories of tasks.

\subsection{Table Extraction} 
In the early years, research was mainly focused on detecting, identifying, and extracting tables from webpages, and classifying them according to some type taxonomy.  Gradually, spreadsheet documents were also considered for table extraction, and type taxonomies became more fine-grained.  With the advancement of table extraction and classification methods, large-scale table corpora were constructed, which became available as resources to be utilized in other tasks.  One open issue is that the available table corpora are all a result of a one-off extraction effort.  As such, these collections get quickly outdated.  

\subsection{Table Interpretation} 
The problem of uncovering table semantics, including but not limited to identifying table column types, linking entities in tables, and extracting relational data from tables, represents an active research area.  It is also an important one, as the resulting semantic annotations are heavily utilized in other table-related tasks, such as knowledge base augmentation and question answering.  While there exist methods for high-precision extraction, there is plenty of room for improvement in terms of recall, as most existing methods can only interpret a small portion of tables. For instance, \citet{Ritze:2016:PPW} find that only 2.85\% of web tables can be matched to DBpedia. 
Further, most of the emphasis has been on relational tables; other table types (e.g., entity tables) bring about a different set of challenges.
Another line of future research concerns the development of user interfaces and tools for facilitating and visualizing the annotations~\cite{Mazumdar:2019:ATC}.

\subsection{Table Search}
The task of retrieving relevant tables from a table corpus for an input query is one of the core tasks that was started in the early days and remains to be an active research topic ever since.  One limitation of existing work is that it often makes assumptions about underlying query intent and the preferred answer table types.   
For example, \citet{Zhang:2018:AHT} assume that queries follow a class-property pattern, which can be successfully answered by relational tables. As a result, relational tables with this pattern are preferred, which might therefore result in lower coverage. 
TableNet~\citep{Fetahu:2019:TAD}, a recent study on the interlinking of tables with subPartOf and equivalent relations, can provide a better understanding of table patterns.
In the future, it would be desirable if an automatic query intent classifier were to identify the type of result table sought, which does not need to be limited to relational tables.
Another topic that deserves attention in our opinion, but has not been explored yet, is the presentation of table search results.  For example, for large tables, how should appropriate snippets (summaries) be generated for search result pages? 
Following the two main lines of approaches to automatic summarization, summaries could be extraction-based, by selecting relevant portions of the table to be displayed, or abstraction-based, by generating a natural language summary of its contents~\cite{Hancock:2019:GTW}.
		
\subsection{Question Answering}
Facts extracted from tabular data can be used for answering natural language questions. Previous work has looked at answering questions using a single table or multiple tables. 
Much of the research emphasis has been directed to parsing the questions and on extracting the facts from tables.
While studying these, certain simplifications were made concerning other aspects of the problem. 
For example, works that address QA on a single table all take a carefully selected table (which is to be treated as a knowledge base) for granted.  Locating a proper KB table is a challenging table search task that remains to be solved.
There also seems to be a lack of understanding of when tables can actually aid QA.  Existing research has found that even though QA on tables suffers from low coverage, it can complement QA on text.  Yet, there has not been any systematic study on understanding what are the types of questions where tables can help or what is the scope of facts or relations where web tables have sufficient coverage.
Finally, the heterogeneity of web tables limits the applicability of current methods to a small portion of tables. In the future, more generic methods would need to be developed to be able to deal with heterogeneous tables.

\subsection{Knowledge Base Augmentation} 
The knowledge mined from tables can (also) be utilized for knowledge base (KB) population/construc\-tion. Existing methods address one of two main problems: (1) matching tables to a knowledge base or (2) discovering new facts/relations from tables by utilizing these ``table-to-KB'' matching results. Yet, all these approaches seem to ignore ``out-of-KB'' data, that is, entities or properties that are not linked to a knowledge base. 
Wikipedia tables are one specific example that contain many unlinked entity mentions. 
Those entities/properties could also be potentially useful for populating KBs with new information. 
Shortcomings of current approaches include (1) the lack of consideration for temporal information and (2) identifying entities at the right level of granularity (e.g., location may be given as a city or as a state or country)~\cite{Ritze:2016:PPW}. The former is especially important, as it may promote further utilization of tables to help keep KBs up-to-date.

\subsection{Table Augmentation}
There is a solid body of work on augmenting existing tables with additional data, extracted either from other tables or from knowledge bases.  However, there are at least two issues that remain. One is tapping into the large volumes of unstructured sources (e.g., webpages). The other is combining data from multiple sources, which brings about a need for techniques to draw users' attention to conflicting information and help them to deal with those cases.
Existing work on augmentation assumes the presence of an input table that the user needs helps with completing. An exciting and ambitious research direction is to automatically generate a table, which can answer the user's information need, from scratch~\citep{Zhang:2018:OTG}.

%\input{00paper-09}

%\bibliographystyle{unsrt}  

\bibliographystyle{ACM-Reference-Format}
\bibliography{00paper}

\end{document}